\documentclass[12pt]{article}
\usepackage{graphicx,amsmath,amssymb}

\numberwithin{equation}{section} 

\begin{document}

\thispagestyle{empty}
\vspace*{.2cm}
\noindent
MPP-2010-174 \hfill DESY 11-015 \\

\vspace*{1.0cm}

\begin{center}

{\bf {\Large Towards Matter Inflation in Heterotic String Theory}}
\\[1.5cm]

{\bf Stefan~Antusch$^{\rm 1}$\let\thefootnote\relax\footnotetext{E-mails:~~antusch@mpp.mpg.de},
Koushik Dutta$^{\rm 2}$\footnote{\hspace*{1.5cm}koushik.dutta@desy.de}, 
Johanna~Erdmenger$^{\rm 1}$\footnote{\hspace*{1.5cm}jke@mpp.mpg.de},
Sebastian~Halter$^{\rm 1}$\footnote{\hspace*{1.5cm}halter@mpp.mpg.de}
\\[3mm]
\normalsize
$^{\rm 1}$ {\it Max-Planck-Institut f\"ur Physik (Werner-Heisenberg-Institut), \\ F\"ohringer Ring 6, D-80805 M\"unchen, Germany} \\
\normalsize
$^{\rm 2}$ {\it Deutsches Elektronen-Synchrotron DESY, \\ Notkestrasse 85, D-22603 Hamburg,
Germany}
}\\[1.5cm]

{\bf Abstract}
\end{center}

\noindent Recently, a class of inflation models in supergravity with gauge non-singlet matter fields as the inflaton has been proposed. It is based on a `tribrid' structure in the superpotential and on a `Heisenberg symmetry' for solving the $\eta$-problem. We suggest that a generalization of this model class may be suitable for realising inflation in heterotic orbifold compactifications, where the Heisenberg symmetry is a property of the tree-level K\"ahler potential of untwisted matter fields. We discuss moduli stabilization in this setup and propose a mechanism to stabilize the modulus associated to the inflaton, which respects the symmetry in the large radius limit. Inflation ends via a ‘waterfall’ phase transition, as in hybrid inflation. We give conditions which have to be satisfied for realising inflation along these lines in the matter sector of heterotic orbifolds.

\newpage

\setcounter{footnote}{0}

\tableofcontents

\section{Introduction}
\label{Sect_Introduction}

Inflation is a successful paradigm for solving the flatness and horizon problems of Standard Big Bang cosmology and for providing the seed of structure in the universe \cite{Lyth:2009zz}. However, there are various open questions, for example: What is the inflaton, i.e.\ the particle responsible for the inflationary dynamics? How can the necessary flatness of the inflaton potential be realised? The latter challenge is known as the $\eta$-problem  \cite{Copeland:1994vg,Dvali:1995mj}, named after the slow-roll parameter $\eta$ which should be substantially smaller than one to obtain consistent slow-roll inflation. The $\eta$-problem is a very generic challenge: In the inflationary era a large vacuum energy $V_0$ contributes dominantly to the energy density of the universe. From the perspective of effective field theory, dimension six operators of the form $\mathcal{O}_{4} \phi^2 / M_{P}^2$ with $\langle \mathcal{O}_4 \rangle \sim V_0$ give a mass to the inflaton field $\phi$ of the order $V_0 / M_{P}^2$, i.e. of the order of the Hubble scale. This would violate the above-mentioned condition for slow-roll inflation due to $\eta \sim 1$.

At the level of effective field theories, for instance in supergravity, one may postulate approximately conserved symmetries like a `shift symmetry' \cite {Kawasaki:2000yn,Brax:2005jv} or a `Heisenberg symmetry' \cite{Stewart:1994ts,Gaillard:1995az} to solve the $\eta$-problem. Recently, some progress has been made with respect to the second possibility: In \cite{Antusch:2008pn} a new class of inflation models in supergravity has been proposed, where the approximately conserved Heisenberg symmetry solves the $\eta$-problem and where the associated modulus is stabilized by corrections to the K\"ahler potential. Inflation ends by the `waterfall mechanism' as in hybrid inflation \cite{Linde:1993cn}. However, compared to standard SUSY hybrid inflation \cite{Dvali:1994ms}, the new class of models dubbed `tribrid inflation' is taylor-made for using symmetry solutions to solve the $\eta$-problem \cite{Stewart:1994ts,Antusch:2009vg} (for some recent related work see also \cite{Kallosh:2010xz}). It has furthermore been demonstrated in \cite{Antusch:2010va} that tribrid inflation combined with the Heisenberg symmetry allows the inflaton to be a (combination of) gauge non-singlet field(s) in the matter sector of the theory.

To address the question of the origin of the symmetries used for solving the $\eta$-problem, it is necessary to go beyond the effective field theory approach and consider a UV-completion of the theory, i.e. a theory of quantum gravity.\footnote{For recent attempts of `low energy solutions' of this issue see \cite{Baumann:2010ys,Baumann:2010nu}.}  The main reason for this is that black hole evaporation suggests that continuous global symmetries with associated conserved charges are broken in the UV \cite{Kallosh:1995hi}. The leading candidate for a UV-completion to date is string theory and various approaches have been followed to realise inflation in string theory. An important class of models is that of brane-anti-brane inflation: The inflaton is in the open string sector and corresponds to the position of a D-brane moving towards an anti-D-brane in the compact space (for early work on brane inflation see \cite{Dvali:1998pa}). This idea is viable as `warped' D-brane inflation \cite{Kachru:2003sx}, where a D3-brane moves down a warped throat, and which has been further developed in \cite{Berg:2004ek}. There are also models of brane inflation with D3-branes moving towards D7-branes, see e.g. \cite{Haack:2008yb}. Another class of models where the inflaton is a modulus from the closed string sector, i.e. the volume of a cycle of the compact space, has been investigated rigorously \cite{BlancoPillado:2004ns}. Very recently, models of large field inflation based on the concept of monodromy have been constructed \cite{Silverstein:2008sg}. For recent reviews and further references on models of inflation in string theory see \cite{McAllister:2007bg}.

The issue of moduli stabilization is crucial for any model of inflation in string theory. Most constructions are in the context of type IIB string theory, more precisely they are based on `(warped) flux compactifications' \cite{Giddings:2001yu}. So far, this class of models provides arguably the best understood framework for moduli stabilization in string theory \cite{Kachru:2003aw}. For reviews and further references on flux compactifications see \cite{Grana:2005jc}.

In this paper, we propose a new approach for realising inflation in the matter sector of heterotic string models. In many orbifold compactifications of the heterotic string the effective four-dimensional action features a Heisenberg symmetry in the untwisted sector \cite{Binetruy:1987xj,Ellwanger:1986yh}. The Heisenberg symmetry is a non-compact symmetry, which extends the shift symmetry of the axion contained in a K\"ahler modulus $T$. It allows the K\"ahler modulus and the matter fields, to be identified as inflaton(s), to appear in the K\"ahler potential only in a certain combination. The symmetry allows to keep the inflaton direction flat at tree-level, thereby alleviating the $\eta-$problem \cite{Gaillard:1995az}. Regarding string inflation with Heisenberg symmetry, it has been suggested in \cite{Gaillard:1998xx} that a phase of vacuum energy dominance can emerge due to the symmetry, but no model for the inflationary dynamics or the end of inflation was proposed. Our approach is based on a generalization of the tribrid inflation class of models where the matter fields enjoy a Heisenberg symmetry. Here, we outline how this scenario can be embedded into heterotic orbifolds with some untwisted matter fields as the inflatons. The conditions on the superpotential in the tribrid setup are well-suited for heterotic orbifolds. Moreover, since we aim at realising inflation in the matter sector of the theory, we wish to choose a class of string compactifications where MSSM-like spectra can be obtained. Indeed, a certain class of heterotic orbifolds, the `mini-landscape' models have been thoroughly studied and shown to yield spectra very close to the MSSM \cite{Buchmuller:2005jr}. The most recent attempts towards (bulk) moduli stabilization in these models are \cite{Dundee:2010sb,Parameswaran:2010ec}, where the latter tries to construct explicit examples. However, there is so far no explicit model with all bulk moduli stabilized. Here, we address inflation in heterotic orbifold compactifications, but we do not focus on a particular class of orbifolds. Instead, we propose a framework which is in principle capable of achieving inflation in the matter sector and stabilizing the bulk moduli during inflation. We describe the necessary conditions for inflation in this setup and it remains to be checked whether they can be satisfied in phenomenologically interesting compactifications.

One important aspect of our scenario is that we expect the K\"ahler modulus associated to the Heisenberg symmetry protecting the inflaton and the dilaton to be stabilized during inflation only by particular terms in the K\"ahler potential originating from perturbative and non-perturbative corrections. Note that this setup is in sharp contrast to the standard picture where moduli stabilization both during and after inflation is achieved by terms in the superpotential, e.g. as in the KKLT scenario for type IIB flux compactifications \cite{Kachru:2003aw}. The standard procedure has the advantage of better control over the responsible terms due to the non-renormalization theorems for the superpotential. However, in these models there is often a tension between high scale inflation and low scale supersymmetry breaking \cite{Kallosh:2004yh}. Typically, the Hubble scale during inflation is constrained to be less than today's gravitino mass in order to avoid decompactification, i.e. $H_{\text{inf}} \lesssim m_{3/2}$. In our case, this problem is relaxed since we employ different mechanisms for moduli stabilization during and after inflation (for example, a gaugino condensate is typically negligible during inflation but can be important afterwards \cite{Gaillard:1998xx}). Therefore, there is no direct relation between the Hubble scale during inflation and the gravitino mass today.\footnote{For a recent proposal to solve this issue in KKLT-type models see \cite{He:2010uk}, where the moduli stabilizing part of the superpotential changes during inflation.}

The dilaton is stabilized during inflation by non-perturbative corrections to the K\"ahler potential as in \cite{Gaillard:1998xx}. We propose a way to stabilize the K\"ahler modulus associated to the inflaton which we expect to break the Heisenberg symmetry only weakly in the large radius limit. It is based on moduli-dependent threshold corrections to the K\"ahler metric of a twisted matter field. If there are sectors with $\mathcal{N} = 2$ supersymmetry, such corrections can be present and there are known results for the K\"ahler metric of untwisted matter fields \cite{Antoniadis:1992pm} (the $\mathcal{N} = 1$ sectors contribute only moduli-independent corrections). We assume that the corrections to the K\"ahler metric of twisted fields take a similar functional form. Moreover, as a working hypothesis, we consider the case where these corrections preserve the Heisenberg symmetry up to terms which are exponentially suppressed in the large radius limit. This conjecture is based on the observation that the shift symmetry protecting the imaginary part of the modulus involved in the threshold corrections is recovered as an approximate symmetry at large radius (it is only broken by worldsheet instantons \cite{Dine:1986zy,Ferrara:1991uz}). We argue that this form of the K\"ahler metric allows for moduli stabilization during inflation with a minimum at large radius. Hence, we have a sufficiently flat inflaton potential if the Heisenberg symmetry is indeed approximately recovered in this limit. We also briefly discuss alternatives for generating a slope for the inflaton such as loops involving the waterfall fields.

This paper is organized as follows: in Sect.~\ref{Sect_GeneralModel} we describe a class of models, which is a generalization of the `tribrid' models of \cite{Antusch:2008pn} with a Heisenberg symmetry. We explain how the moduli get stabilized during inflation and comment on the inflaton slope and the hybrid mechanism. In Sect.~\ref{Sect_EffectiveAction}, we review the basic ingredients of the effective action of heterotic orbifolds. Based on the ingredients described in Sect.~\ref{Sect_EffectiveAction}, we outline in Sect.~\ref{Sect_Realization} a proposal how to embed the general class of models of Sect.~\ref{Sect_GeneralModel} into heterotic orbifolds. Moduli stabilization during inflation is described and we briefly comment on constraints from imposing D-flatness and the slope of the inflaton potential. We then conclude and summarize open questions in Sect.~\ref{Sect_Conclusions}.

\section{General class of models}
\label{Sect_GeneralModel}

We consider a generalization of the `tribrid' inflation models of \cite{Antusch:2008pn}, which employed a Heisenberg symmetry \cite{Binetruy:1987xj} and a specific structure of the superpotential to solve the $\eta$-problem and implement  the waterfall mechanism of hybrid inflation \cite{Linde:1993cn}. The generalization is done keeping in mind what might be viable in heterotic orbifold compactifications. It is characterized by the following requirements:
\begin{itemize}
	\item A D-flat and F-flat direction of (matter) fields acts as the inflaton.
	\item The inflaton direction is protected against the $\eta$-problem by an approximate symmetry.
	\item The inflaton is coupled to a waterfall sector such that the inflationary part of the superpotential approximately vanishes during (and after) inflation, i.e. $\langle W \rangle \simeq 0$.
\end{itemize}
The condition of an approximately vanishing superpotential helps to suppress higher order
supergravity corrections that usually destroy the flatness of the inflaton potential \cite{Stewart:1994ts}. In addition, this requirement reduces the couplings between the inflaton and other sectors in the theory, e.g a moduli sector. 

First, in Sect.~\ref{Sect_BasicSetup} we review the `tribrid' structure of the superpotential and the Heisenberg symmetry as described in \cite{Antusch:2008pn,Antusch:2009vg}. Then we extend this in Sect.~\ref{Sect_Generalization} to a form which is suitable for a possible realisation of the setup within heterotic orbifolds. We discuss moduli stabilization in Sect.~\ref{Sect_GeneralModuli} and comment on the inflaton slope and the hybrid mechanism in Sect.~\ref{Sect_GeneralSlope}.

\subsection{Basic setup}
\label{Sect_BasicSetup}
In the `tribrid' inflation models, the superpotential is supposed to have the following structure \cite{Antusch:2008pn,Antusch:2009vg}
\begin{equation}
  \label{Eq_GeneralTribridSuper}
  W = \kappa X ( H^2 - M^2 ) + f(\Phi) H^2 \, ,
\end{equation}
where the three fields $X$ $,H$ and $\Phi$ play different roles: they provide the large vacuum energy, the mechanism to end inflation and the `clock' determining when inflation ends (i.e. when $\Phi$ reaches a critical value, one of the waterfall degrees of freedom becomes tachyonic), respectively. The scale $M$ sets the expectation value of $H$ at the end of inflation and the F-term of $X$ during inflation and thus the vacuum energy. This superpotential is supplemented by a K\"ahler potential of the form
\begin{equation}
  \label{Eq_GeneralTribridKaehler}
  K = ( \lvert X \rvert^2 + \lvert H \rvert^2 - \kappa_{X} \lvert X \rvert^4 + \kappa_{XH} \lvert X \rvert^2 \lvert H \rvert^2 + \dots ) + d(\rho) \lvert X \rvert^2 + k(\rho) \, ,
\end{equation}
where $\rho$ contains the inflaton $\Phi$ and a modulus $T$ in a combination which is invariant under the Heisenberg symmetry:
\begin{equation}
  \label{Eq_GeneralTribridx}
  \rho = T + \bar{T} - \lvert \Phi \rvert^2 \, .
\end{equation}
Examples for the functions $d(\rho)$ and $k(\rho)$ can be found below and in \cite{Antusch:2008pn}. The Heisenberg symmetry acting on $T$ and $\Phi$ consists of the following two elements \cite{Binetruy:1987xj}:
\begin{equation}
  \label{Eq_GeneralHeisenberg1}
  T \rightarrow T + i \alpha \quad , \quad \alpha \in \mathbb{R} \, ,
\end{equation}
and
\begin{equation}
  \label{Eq_GeneralHeisenberg2}
  \begin{split}
    T & \rightarrow T + \bar{\beta} \Phi + \frac{1}{2} \bar{\beta} \beta \, , \\
    \Phi & \rightarrow \Phi + \beta \quad , \quad \beta \in \mathbb{C} \, . \\
  \end{split}
\end{equation}
Note that we use units where $M_P = 1$ throughout this work, except where stated otherwise. The K\"ahler potential in Eq.~\eqref{Eq_GeneralTribridKaehler} is expanded in $H$ and $X$. The $f(\Phi) H^2$ term in the superpotential provides a positive $\Phi$-dependent mass squared for the waterfall field $H$. A possible choice for $f(\Phi)$ is for example $f(\Phi) = \Phi^n$ for any $n \geq 1$. The field $X$ provides the vacuum energy by its F-term and the function $k(\rho)$, in combination with a suitable choice of $d(\rho)$, can stabilize $\rho$ during inflation. One main feature of this framework is that during inflation we have $\langle W \rangle = 0, \langle W_{\Phi} \rangle = \langle W_{H} \rangle =  \langle W_{T} \rangle = 0$, $\langle W_{X} \rangle \neq 0$ with $\langle X \rangle = \langle H \rangle = 0$. It was emphasized in \cite{Stewart:1994ts, Antusch:2008pn} that these conditions are desirable for solving the $\eta$-problem with a Heisenberg symmetry. Because of the symmetry, the inflaton direction is protected from the $\eta-$problem, whereas the symmetry breaking term $f(\Phi) H^2$ in the superpotential provides the required slope via quantum loop corrections to the tree-level flat potential. 

One interesting aspect of the Heisenberg symmetry is that it allows the inflaton $\Phi$ to be a gauge non-singlet matter field. We allow for this below, following \cite{Antusch:2010va}, where an explicit example of matter inflation in the context of supersymmetric Grand Unified Theories (GUTs) was constructed. This requires a modification of Eqs.~\eqref{Eq_GeneralTribridSuper} and \eqref{Eq_GeneralTribridKaehler}. In particular, one has to introduce multiple matter fields $\Phi_a$ in order to satisfy the constraints from imposing D-flatness. Then the function $f(\Phi)$ in Eq.~\eqref{Eq_GeneralTribridSuper} has to involve a gauge-invariant product of these matter fields $\Phi_a$, for example $f(\Phi_a) = \Phi^{+} \Phi^{-}$. In addition, the combination $\rho$ in Eq.~\eqref{Eq_GeneralTribridx} gets modified to $\rho = T + \bar{T} - \sum_{a} \lvert \Phi_a \rvert^2$. Moreover, the field $H$ is replaced by two matter fields $H^{\pm}$ in conjugate representations, i.e. in all terms in the superpotential $H^2$ now becomes $H^{+} H^{-}$. Note also that now the parameter $\beta$ in the symmetry transformation in Eq.~\eqref{Eq_GeneralHeisenberg2} has to be replaced by a set of parameters $\beta_a$. The Heisenberg symmetry is an approximate symmetry in the limit of vanishing superpotential and gauge coupling. Thus, we must ensure $\langle W \rangle \simeq 0$ during inflation and that there are no background gauge fields under which the inflaton is charged. The corrections induced by loops involving gauge fields are discussed in Sect.~\ref{Sect_GeneralSlope}.

\subsection{A generalization}
\label{Sect_Generalization}
The superpotentials and K\"ahler potentials we consider from now on are a further generalization of the `tribrid' structure that is introduced in the last section:
\begin{subequations}
  \label{Eq_GeneralModel}
  \begin{align}
  \label{Eq_GeneralSuperpotential}
  W = & a(T_i) \, X \left[ b(T_i) \, H^{+} H^{-} - \langle \Sigma \rangle^{2} \right] + c\,(T_i)  \, f(\Phi_a) \, H^{+} H^{-} + \widetilde{W} \, , \\
  \label{Eq_GeneralKaehlerpotential}
  K = & - \log(T_1 + \bar{T}_1) - \log(T_2 + \bar{T}_2) - \log \left(T_3 + \bar{T}_3 - \sum_{a} \lvert \Phi_a \rvert^2 \right)  + \widetilde{K} \, .
  \end{align}
\end{subequations}
The first three terms in the K\"ahler potential are the analog of $k(\rho)$ in Eq.~\eqref{Eq_GeneralTribridKaehler}, and $\widetilde{K}$ includes the analogs of all the other terms in Eq.~\eqref{Eq_GeneralTribridKaehler}. The expectation value $\langle \Sigma \rangle$ replaces the mass scale $M$ in Eq.~\eqref{Eq_GeneralTribridSuper}. We will now first describe the field content of the model and then the structure of the superpotential and K\"ahler potential in more detail.

The $T_i$, $i = 1, 2, 3$, are moduli fields and their K\"ahler potential satisfies the no-scale property (i.e. $K_i K_{\bar{j}} K^{i \bar{j}} = 3$ for $i, j = T_1, T_2, T_3$). The $\Phi_a$ denote matter fields, i.e. gauge non-singlet fields, and $f(\Phi_a)$ is supposed to be a gauge-invariant product of the $\Phi_a$, such that a D-flat combination of these fields can act as the inflaton. That is, the inflaton is a certain linear combination of the $\Phi_a$, which is specified by the vanishing of the D-terms.  Such a gauge-invariant product may be as simple as $\Phi^+ \Phi^-$, but it can also be a more general combination. The Heisenberg symmetry demands that the K\"ahler potential depends only on the invariant quantity
\begin{equation}
  \label{Eq_GeneralInvariantrho3}
  \rho_3 \equiv T_3 + \bar{T}_3 - \sum_{a} \lvert \Phi_a \rvert^2 \, .
\end{equation}
The choice to associate the $\Phi_a$ to $T_3$ is of course arbitrary. The waterfall fields $H^{\pm}$ belong to conjugate representations with respect to some gauge group, e.g. a $U(1)$, as is indicated by the superscript $\pm$. They are kept at zero during inflation due to the $f(\Phi_a) H^+ H^-$ term in $W$ for sufficiently large values of the $\Phi_a$ and acquire expectation values at the end of inflation. Since the $\Phi_a$ are gauge non-singlets, $f(\Phi_a)$ is at least quadratic in the fields and thus involves a scale $\Lambda$, e.g. $f(\Phi_a) = \Phi^{+} \Phi^{-} / \Lambda$. This scale is a priori undetermined. In the case of heterotic orbifolds considered later we will take $\Lambda \sim M_{s}$, where $M_{s}$ denotes the string scale.

The field $X$ receives an F-term during inflation (where $\langle H^{\pm} \rangle = 0$) and thereby provides the vacuum energy. Moreover, its expectation value is fixed at $\langle X \rangle = 0$ during inflation, e.g. enforced by a term $- \kappa_{X} \, \lvert X \rvert^4$ with $\kappa_{X} > 0$ and sufficiently large in the K\"ahler potential \cite{Kawasaki:2000yn}.\footnote{From the general supergravity analysis of \cite{Covi:2008cn} one can give a geometric interpretation to requiring a negative $\lvert X \rvert^4$ term in the K\"ahler potential: What matters is the curvature of the scalar manifold along the supersymmetry breaking direction, here $X$, which has to be negative in the limit $W \rightarrow 0$, i.e. for vanishing gravitino mass $m_{3/2}$.} At the end of inflation, its F-term vanishes once the waterfall fields $H^\pm$ acquire their expectation values, which also ensure $\langle X \rangle = 0$ after inflation. The scale of the F-term of $X$ is assumed to be generated by the expectation value of a collection of fields denoted by $\langle \Sigma \rangle$, which is e.g. induced by the Fayet-Iliopoulos D-term of an anomalous $U(1)_{A}$.

An example for a possible D-flat trajectory suitable for inflation is
\begin{equation}
    \label{Eq_GeneralDflatdirection}
    \langle X \rangle = \langle H^\pm \rangle = 0 \,\, , \,\, \lvert \langle \Phi^+ \rangle \rvert = \lvert \langle \Phi^- \rangle \rvert \, ,
\end{equation}
if $f(\Phi_a)$ is chosen to be $\Phi^{+} \Phi^{-} / \Lambda$. The D-term equation relates the absolute values of $\Phi^+$ and $\Phi^-$ while their phases remain undetermined. We identify the inflaton with the unfixed absolute value. In principle, $f(\Phi_a)$ can have a more general form and then the D-term equations become more complicated.

The function $a(T_i)$ must depend only on $T_1$ and $T_2$ in order not to spoil the flatness of the potential at the tree level. The functions $b(T_i)$ and $c(T_i)$ on the other hand may depend on all three moduli since $\langle H^{\pm} \rangle = 0$ during inflation.

The terms in the superpotential which are subleading with respect to the F-term of $X$ are collectively denoted by $\widetilde{W}$. In particular, those terms may be responsible for moduli stabilization after inflation and also may be important for low-energy supersymmetry breaking. Here, since supersymmetry breaking during inflation by the F-term of $X$, we can work in the approximation $W_X \neq 0$, $W \simeq W_n \simeq 0$, where the index $n$ runs over all fields other than $X$.

The function $\widetilde{K}$ determines in particular the kinetic terms of $H^\pm$ and $X$ and furthermore includes all other terms not relevant for our inflation setup. Since we keep $X$ and $H^\pm$ at zero during inflation, let us expand the K\"ahler potential in powers of $X, H^\pm$. To quadratic order in $\Upsilon_{\alpha} \in \{ X, H^\pm \}$, $\widetilde{K}$ is of the form
\begin{equation}
  \label{Eq_GeneralXkinetic1}
  \widetilde{K} = k_{\alpha}(T_1 + \bar{T}_1, T_2 + \bar{T_2}, \rho_3) \lvert \Upsilon_{\alpha} \rvert^2 + \dots \, ,
\end{equation}
where the dots denote higher order terms in $\Upsilon_{\alpha}$ as well as terms independent of $\Upsilon_{\alpha}$. We assume that the moduli-dependent functions $k_{\alpha}$ respect the Heisenberg symmetry and consider two possible functional forms, which are motivated by heterotic string compactifications:
\begin{equation}
  \label{Eq_GeneralXkinetic2}
  k_{\alpha} = (T_1 + \bar{T}_1)^{- q_{1}^{\alpha}} \, (T_2 + \bar{T}_2)^{- q_{2}^{\alpha}} \, \rho_{3}^{- q_{3}^{\alpha}}  \, ,
\end{equation}
for $\alpha = H^\pm$ and for $\alpha = X$
\begin{equation}
  \label{Eq_GeneralXkinetic3}
  k_X = \frac{1 + d(\rho_3)}{\left( T_1 + \bar{T}_1 \right)^{q_{1}^X} \left( T_2 + \bar{T}_2 \right)^{q_{2}^X}} \, .
\end{equation}
The $q_{i}^{\alpha} \geq 0$, $i = 1, 2, 3$, are model-dependent rational numbers (with $q_{3}^{X} = 0$).

Using $W \simeq W_n \simeq 0$ for $n \neq X$, the F-term potential during inflation (where $X = H^{\pm} = 0$) is given by
\begin{equation}
  \label{Eq_GeneralFtermpotential}
  V_F \simeq e^{K} \left( K_{X \bar{X}} \right)^{-1} \lvert W_X \rvert^2 = \frac{e^{K} \lvert a(T_1, T_2) \rvert^2 \lvert \langle \Sigma \rangle \rvert^4}{k_{X}(T_1 + \bar{T}_1, T_2 + \bar{T}_2, \rho_3)} \, ,
\end{equation}
where $k_X$ is given by Eq.~\eqref{Eq_GeneralXkinetic3}, $K$ given by Eq.~\eqref{Eq_GeneralKaehlerpotential} and the expectation value of $\Sigma$ can be induced either through the D-term of an anomalous $U(1)$ or through the superpotential. The role of $\Sigma$ is to set the scale of the F-term of $X$ through its expectation value, and thus the overall scale of the scalar potential.

\subsection{Moduli stabilization during inflation}
\label{Sect_GeneralModuli}

The moduli fields $T_i$ have to be stabilized during inflation. Since none of them is the inflaton, we would like to stabilize them with a high mass at least of the order of the Hubble scale $H_{\text{inf}}$. For $T_1$ and $T_2$, stabilization is achieved by a suitable form of the function $a(T_1,T_2)$ in Eq.~\eqref{Eq_GeneralSuperpotential}, which enters the F-term of $X$. Moreover, the modulus $T_3$, or rather the combination $\rho_3$ in Eq.~\eqref{Eq_GeneralInvariantrho3}, is fixed by an appropriate moduli dependence of the kinetic term of $X$, i.e. by an appropriate choice of $d(\rho_3)$. We now discuss how both stabilization mechanisms work in a simple toy model. Note that due to the product structure of the moduli dependence in Eq.~\eqref{Eq_GeneralFtermpotential} we can discuss the stabilization of $T_1, T_2$ and $\rho_3$ separately.

\begin{figure}[ht]
\begin{center}
\includegraphics[scale=0.8]{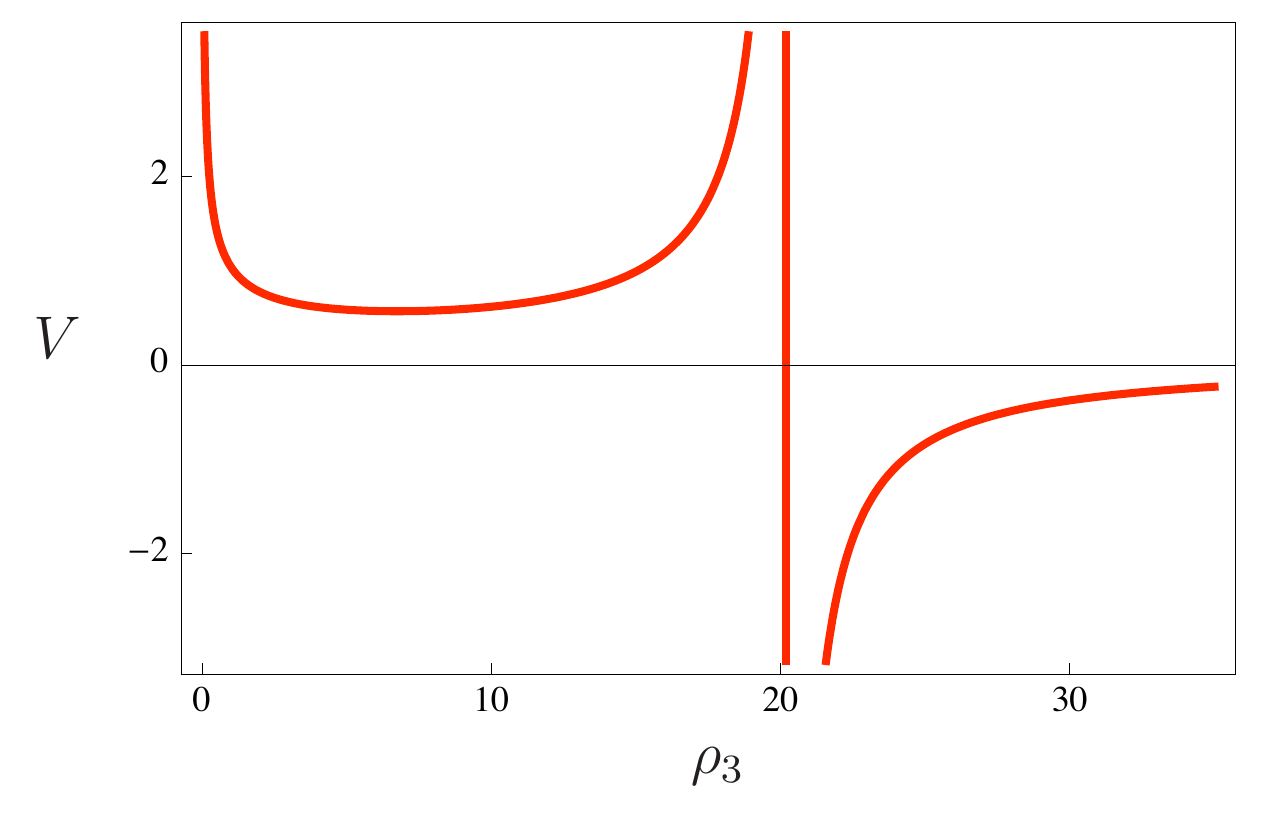}
\caption{Form of the potential following from Eq.~\eqref{Eq_GeneralPotentialrho3} for the example values $p = - \frac{1}{2}$, $\gamma = 0.01$ and $\beta = -0.05$, which has a minimum at $\langle \rho_3 \rangle \approx 6.73 $. The pole is at $\rho_3 = 20.2$. The overall scale of the potential has to be set by $\langle \Sigma \rangle$.}
\label{Fig_GeneralPotentialrho3}
\end{center}
\end{figure}

We stabilize $\rho_3$ during inflation with the F-term of $X$ combined with a suitable kinetic term for $X$ in order to give it a mass $m \gtrsim H_{\text{inf}}$. The moduli dependence of the F-term potential Eq.~\eqref{Eq_GeneralFtermpotential} not only depends on $K$, $k_X$ and $a(T_i)$ but also on $\langle \Sigma \rangle$. As we will see in Sect.~\ref{Sect_AnomalousU1andVEVs}, due to the moduli-dependent non-canonical K\"ahler potential (and superpotential) terms, the expectation value $\langle \Sigma \rangle$ typically inherits some moduli dependence. Assuming that the moduli dependence is inherited from the K\"ahler potential, $\langle \Sigma \rangle \propto \rho_{3}^q$ for some rational number $q \geq 0$. The function $d(\rho_3)$ needs to respect the Heisenberg symmetry to a sufficient amount (in order to ensure a sufficiently small inflaton mass) and must be of a suitable form to stabilize $\rho_3$ during inflation.  To illustrate what `suitable form' means, consider the $\rho_3$-dependence of the F-term potential, which is of the form\footnote{If $\langle \Sigma \rangle$ is independent of $\rho_3$, we have $p = -1$. Otherwise it is some rational number.}
\begin{equation}
  \label{Eq_GeneralPotentialrho3}
  V \propto  \frac{\rho_{3}^p}{1 + d(\rho_3)} \, ,
\end{equation}
with some rational number $p \geq -1$. In order to get a minimum suitable for inflation, let us make the simple ansatz $d(\rho_3) = \gamma + \beta \rho_3$. If $p < 0$, $\beta < 0$ and $\gamma > -1$, this yields a minimum at
\begin{equation}
  \label{Eq_GeneralMinimumValuerho3}
  \langle \rho_3 \rangle = - \frac{p \, (1 + \gamma)}{(p - 1) \, \beta} > 0 \, .
\end{equation}
For this choice of $d(\rho_3)$, the potential has a pole at $\rho_3 = - (1 + \gamma) / \beta$, which provides the barrier towards $\rho_3 \rightarrow \infty$. For $p < 0$, the $\rho_{3}^p$ factor prevents the field from rolling to $\rho_3 \rightarrow 0$. Therefore, the parameter $p$ is constrained to be $ -1 \leq p < 0$. Fig.~\ref{Fig_GeneralPotentialrho3} shows a plot of the $\rho_3$-dependence of the potential (in arbitrary units) for an illustrative choice of parameters.

We expect the pole in the potential to be an artefact of our approximation: we work at second order in the derivatives and the pole appears when $K_{X \bar{X}} \rightarrow 0$ (recall that $V \propto (K_{X \bar{X}})^{-1}$). Therefore, this approximation breaks down close to the pole and higher derivative corrections become important. In particular, if one would like to address issues such as stability of the minimum with respect to tunneling, the higher derivative terms have to be included. Note that in the region to the right of the pole, $K_{X \bar{X}} < 0$ and thus $X$ has a kinetic term with the wrong sign. For our present purpose we only need that the potential given by Eq.~\eqref{Eq_GeneralPotentialrho3} is a good approximation if we are not too close to the pole and we will assume that we are confined within this region.

To determine the physical mass of $\rho_3$ around its minimum in units of the Hubble scale $H_{inf}$ during inflation, we have to take into account that the K\"ahler potential Eq.~\eqref{Eq_GeneralKaehlerpotential} leads to a non-canonical kinetic term for $\rho_3$, namely $\rho_3^{-2}(\partial_{\mu} \rho_3)^2$. Hence, the physical mass is given by 
\begin{equation}
m_{\rho_3}^2 = 2 p (p-1) V_0 \, ,
\end{equation} 
where $V_{0}$ denotes the value of the potential at the minimum (recall that we have set $M_P = 1$ and thus $H_{\text{inf}}^2 \sim V_0$). For $p \lesssim - 0.15$, we have $m_{\rho_3} \gtrsim H_{\text{inf}}$, which should be heavy enough such that $\rho_3$ settles to its minimum sufficiently fast.

\begin{figure}[ht]
\begin{center}
\includegraphics[scale=0.8]{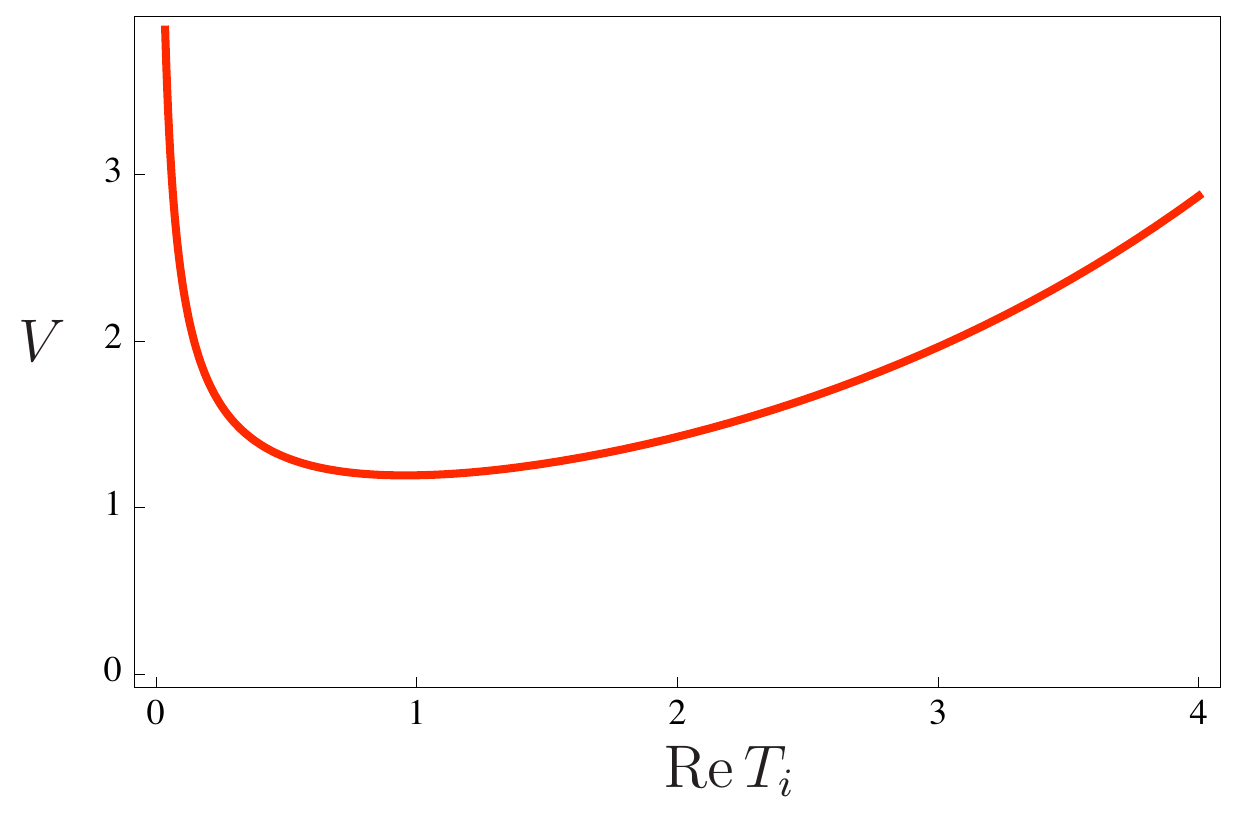}
\caption{Form of the potential following from Eq.~\eqref{Eq_GeneralPotentialT1T2} with respect to $T_i$ for the example values $p_i = - \frac{1}{2}$ and $a_i = \frac{\pi}{12} \approx 0.26 $, which has a minimum at $\langle  \text{Re} T_i \rangle \approx  0.96$. The overall scale of the potential has to be set by $\langle \Sigma \rangle$.}
\label{Fig_GeneralPotentialT1}
\end{center}
\end{figure}

As noted above, stabilization of $T_1$ and $T_2$ requires a `suitably chosen' function $a(T_1,T_2)$. Assuming again that a possible moduli-dependence of $\langle \Sigma \rangle$ is only due to the non-canonical K\"ahler metric, the dependence of the scalar potential on $T_1$ and $T_2$ takes the form\footnote{If $\langle \Sigma \rangle$ is independent of $T_i$, we have $p_i = -1 + q_{i}^{X}$ and otherwise it is some other rational number.}
\begin{equation}
  \label{Eq_GeneralPotentialT1T2}
  V \propto \left( T_1 + \bar{T}_1 \right)^{p_{1}} \left( T_2 + \bar{T}_2 \right)^{p_{2}} \, \lvert a(T_1,T_2) \rvert^2 \, ,
\end{equation}
with $p_1$ and $p_2$ rational numbers $\geq -1$. A simple choice for $a(T_1,T_2)$ which does the job is $a(T_1,T_2) = e^{a_1 T_1 + a_2 T_2}$. If $a_i > 0$ and $p_i < 0$, this will yield a minimum for $\text{Re} \, T_1$ and $\text{Re} \, T_2$: The exponentials diverge as $\text{Re} \, T_i \rightarrow \infty$ and similarly the power law factors diverge as $\text{Re} \, T_i \rightarrow 0$. The minima are at $\langle \text{Re} \, T_i \rangle = - \frac{p_i}{\sqrt 2 a_i} > 0$, which is typically $\mathcal{O}(1)$. Again, taking into account canonical normalization, also $T_1$ and $T_2$ are stabilized with masses $m \sim H_{\text{inf}}$. A plot of the $T_i$-dependence of the potential (in arbitrary units) for a sample choice of parameters is shown in Fig.~\ref{Fig_GeneralPotentialT1}.

Note that in both situations, Eq.~\eqref{Eq_GeneralPotentialrho3} and Eq.~\eqref{Eq_GeneralPotentialT1T2}, for our choices of $d(\rho_3)$ and $a(T_1, T_2)$ the axions associated to $\text{Im} \, T_i$ only receive a potential from the terms contained in $\widetilde{W}$. However, this is not a problem for the inflationary scenario discussed here: The axions are effectively frozen during inflation due to the strong Hubble damping \cite{Antusch:2008pn}.

Once the moduli have settled to their minima, the functions $a, b$ and $c$ can be effectively treated as constants. Similarly, the non-canonical kinetic terms approximately amount to a rescaling of the fields by a constant only.

Actually, the choices for $a(T_1, T_2)$ and $d(\rho_3)$ in this section are not made purely for illustrative purposes, but are motivated from what we expect to find in heterotic orbifold compactifications, as we discuss below in Sects.~\ref{Sect_ModularInvariance} and \ref{Sect_LoopCorrectionsKaehler}, respectively.

Note that after inflation a different mechanism for moduli stabilization is required, which contribute to $\widetilde{W}$. The basic idea is that moduli stabilization after inflation is achieved by different means, which are encoded in $\widetilde{W}$ and stabilize the moduli at a different scale such that the corrections induced by $\widetilde{W}$ do not introduce an $\eta$-problem. We do not discuss this here in detail and leave this issue for future work.

\subsection{A slope for the inflaton and the hybrid mechanism}
\label{Sect_GeneralSlope}

So far, we have generated a large vacuum energy and stabilized all the moduli during inflation. But if the Heisenberg symmetry was exactly preserved, there would be no slope for the inflaton and hence no way to end inflation. In this section, we briefly discuss what sources could generate a slope within our setup and how we expect inflation to end via the hybrid mechanism.

A slope can be induced by various types of sources, for example
\begin{itemize}
  \item by one-loop Coleman-Weinberg corrections induced by the explicit breaking term $f(\Phi_a) H^{+} H^{-}$ in the superpotential, which couples the inflaton to the waterfall sector,
  \item through loops involving gauge fields,
  \item through the subleading terms in $\widetilde{W}$,
  \item or via Heisenberg symmetry breaking terms in the K\"ahler potential.
\end{itemize}
In the latter two cases, the corresponding terms in the scalar potential must be parametrically small compared to the contribution from the F-term of $X$ and hence yield $\eta \ll 1$. If the Coleman-Weinberg corrections dominate, we expect similar inflationary dynamics as those discussed in \cite{Antusch:2010va}. In either case, once the inflaton reaches a critical value, one of the waterfall fields becomes tachyonic and triggers the waterfall phase transition, thereby ending inflation. 

During the phase transition, topological defects such as cosmic strings could be formed since the waterfall fields are charged under a gauge symmetry e.g. a $U(1)$. Without a specific model it is difficult to decide whether these are problematic or not. First, if the symmetry is broken also during inflation, i.e. if the inflaton is also charged under this gauge symmetry, we expect that as in \cite{Antusch:2010va} topological defects could be avoided due to corrections to the inflaton potential, which lift the degeneracy. Second, the analysis of \cite{Jeannerot:2005mc}, who considered the somewhat similar scenario of `standard' F-term hybrid inflation, finds that the consistency of cosmic strings  with WMAP data depends not only on the value of the gauge coupling but also of a parameter $\kappa$. This parameter is related to the inflationary superpotential used in standard F-term hybrid inflation $W = \kappa \, \Phi \, (H^{+} H^{-} - M^2)$, where $\Phi$ is the gauge singlet inflaton field. Values up to $\kappa \lesssim 10^{-2}$ seem to be consistent with the bounds and allowing for a small contribution from cosmic strings to the power spectrum could also improve the fit to the data. Thus, the issue of topological defects should only be addressed in a specific model and therefore is beyond our present scope.

We now comment on the corrections to the inflaton potential from taking into account loops involving gauge bosons and gauginos following the discussion \cite{Antusch:2010va}, where the one-loop and two-loop corrections to the inflaton mass have been computed in a specific model. The one-loop Coleman-Weinberg potential is given by
\begin{equation}
  \label{Eq_GeneralCWpotential}
  V_{\text{1-loop}} = \frac{1}{64 \pi^2} \, \text{STr} \, \left[  \mathcal{M}^{4}(\Phi_a) \left( \log \left( \frac{\mathcal{M}^{2}(\Phi_a)}{Q^2} \right) - \frac{3}{2} \right) \right] \, ,
\end{equation}
where $Q$ is a renormalization scale and $\text{STr}$ denotes the supertrace, which is taken over all bosonic and fermionic degrees of freedom. We are interested in the $\Phi_a$ dependence of all the masses since this dependence can induce a slope for the inflaton. In addition to the waterfall sector, the gauge sector contributes to the one-loop effective potential since the inflaton is a gauge non-singlet combination. Its expectation value induces masses for some of the gauge fields. However, only sectors with a mass splitting between the bosonic and fermionic degrees of freedom contribute to the supertrace in Eq.~\eqref{Eq_GeneralCWpotential}. The waterfall sector has such a mass splitting, while the gauge sector has no mass splitting if direct supergravity gaugino masses are absent. The supergravity gaugino masses are given by
\begin{equation}
  \label{Eq_GeneralSUGRAGauginoMass}
  \mathcal{L}_{\text{gaugino}} = \frac{1}{4} e^{\langle G \rangle / 2} \left\langle G_{i} \left( G^{-1} \right)^{i \bar{j}} \frac{\partial \bar{f}_{a b}}{\partial \bar{\phi}^{\bar{j}}}  \right\rangle + h.c. \, ,
\end{equation}
where $G = K + \log \lvert W \rvert^2$, $f_{a b}$ denotes the gauge kinetic function and $a, b$ label different gauge groups while $i, j$ label different scalar fields. In our case, during inflation we have $\langle X \rangle \simeq 0$, $\langle W \rangle \simeq 0$ and only $\langle W_X \rangle \neq 0$. Thus, the gravitino mass $m_{3/2} \sim e^{\langle G \rangle / 2} \sim e^{K / 2} \langle \lvert W \rvert \rangle \simeq 0$, which already suppresses most of the contributions to the gaugino masses in Eq.~\eqref{Eq_GeneralSUGRAGauginoMass}. Since $\langle X \rangle \simeq 0$ and only $\langle W_X \rangle \neq 0$, the only contribution which survives in the limit $W \rightarrow 0$ (i.e. which is not suppressed due to the small gravitino mass) vanishes if we assume
\begin{equation}
  \label{Eq_GeneralSUGRAGauginoMassCond}
  \left\langle \frac{\partial \bar{f}_{a b}}{\partial \bar{X}} \right\rangle= 0 \, .
\end{equation}
More precisely, we only have to require that this expectation value does not depend on the inflaton, i.e. there should be no terms such as $X f(\Phi_a)$ contained in $f_{a b}$. Note that we also have to forbid this kind of terms in the superpotential. This can be achieved for example by discrete symmetries which either forbid $X f(\Phi_a)$ at all or force it to appear only together with some additional field(s) whose expectation value(s) vanish. Thus, the corrections from the gauge sector at one-loop are expected to be under control (they are essentially controlled by the small value of the gravitino mass $m_{3/2} \propto \lvert \langle W \rangle \rvert$).

There are potentially dangerous corrections at the two-loop level \cite{Dvali:1995fb}. In \cite{Antusch:2010va}, it was shown that in the large gauge boson mass limit the various two-loop diagrams are suppressed by a universal factor $\frac{\kappa^2}{(4 \pi)^4}$, where in our case $\kappa \equiv a(\langle T_i \rangle) \, b(\langle T_i \rangle)$. Thus, for $\kappa \ll 1$ we expect only negligible two-loop contributions.

\section{Effective action of heterotic orbifolds}
\label{Sect_EffectiveAction}

The main goal of this section is to discuss how the elements of the general scenario of Sect.~\ref{Sect_Generalization} could arise in heterotic orbifold compactifications. We then address the issue of realising this scenario below in Sect.~\ref{Sect_Realization}. Here, we discuss the basic ingredients of the effective supergravity action describing orbifolds of the heterotic string. First, we describe the field content and the structure of the tree-level K\"ahler potential as well as the appearance of the Heisenberg symmetry in Sect.~\ref{Sect_HeteroticOrbifolds}. Then, we briefly discuss target space modular invariance and the constraints it imposes on the superpotential in Sect.~\ref{Sect_ModularInvariance}. Perturbative string loop corrections to the gauge kinetic function and the Green-Schwarz counterterm are the subject of Sect.~\ref{Sect_LoopCorrectionsGaugeKinetic}. In Sect.~\ref{Sect_LoopCorrectionsKaehler}, we first review the known result for the string loop corrections to the K\"ahler metric of matter fields. Then we move on to suggesting a generalization in the presence of background values for matter fields. The corrections from non-perturbative effects, which are crucial to stabilize the dilaton, are introduced in Sect.~\ref{Sect_NonPerturbativeCorrections}. Finally, Sect.~\ref{Sect_AnomalousU1andVEVs} discusses how to generate expectation values for fields, which affect the moduli dependence of the scalar potential.

\subsection{Heterotic orbifolds and Heisenberg symmetry}
\label{Sect_HeteroticOrbifolds}

We start with discussing the field content and the tree-level K\"ahler potential of heterotic orbifolds, where the matter fields can be divided into two categories: untwisted and twisted fields. The Heisenberg symmetry appears in the tree-level K\"ahler potential of the untwisted matter fields and is needed to guarantee a sufficiently flat inflaton potential. The twisted sector fields may serve as candidates for identifying the other fields $X, H^{\pm}$ and $\Sigma$ as (products of) twisted matter fields. Moreover, in any heterotic string compactification there is at least one additional modulus as compared to Sect.~\ref{Sect_Generalization}, the dilaton, which controls the size of the string coupling and therefore also of the gauge coupling.

In orbifold compactifications\footnote{For an excellent review on orbifold compactifications see \cite{Bailin:1999nk}.} of the heterotic string, the six internal directions are compactified on a torus $T^6$ modulo a discrete symmetry group, e.g. a $\mathbb{Z}_N$ group. The compact dimensions can be organized into three complex coordinates:
\begin{equation}
  \label{Eq_OrbifoldComplexCoordinates}
    Z_1  \equiv X_4 + i X_5 \, , \,\, Z_2  \equiv X_6 + i X_7 \, , \,\, Z_3  \equiv X_8 + i X_9 \, .
\end{equation}
The orbifold is characterized by a three dimensional `twist' vector $v$, which encodes the twist acting on the coordinates $Z_i$ as $Z_i \rightarrow e^{2 \pi i v_i} Z_i$ for $i = 1, 2, 3$. For example, the heterotic `mini-landscape' models \cite{Buchmuller:2005jr} based on $\mathbb{Z}_{6-II}$ have the twist vector $v = \frac{1}{6} (1, 2, -3)$, i.e. a rotation by $(60^\circ, 120^\circ, 180^\circ)$ of the first, second and third torus, respectively. The vector $v$ defines the first twisted sector of the theory, and the $k$-th twisted sector is defined by the twist vector
\begin{equation}
  \label{Eq_OrbifoldTwistVector1}
  \eta_{i}(k) \equiv k \, v_i \quad \text{mod} \, 1 \, ,
\end{equation}
where $0 \leq \eta_{i}(k) < 1$ and $k = 1, \dots, N - 1$ for $\mathbb{Z}_N$ orbifolds and one requires in addition
\begin{equation}
  \label{Eq_OrbifoldTwistVector2}
  \sum_{i} \eta_{i}(k) \equiv 1 \, .
\end{equation}
The field content of heterotic orbifolds is therefore devided into two classes: untwisted and twisted sector fields. Geometrically this classification distinguishes fields propagating in all 10 dimensions from those propagating only in 6 or 4 dimensions. The latter are the two types of twisted fields which can arise: they can be either confined to a fixed torus or to a fixed point, depending on the particular twisted sector, i.e. whether the twist leaves one torus unrotated or rotates all three of them. Note that these sectors also have a different amount of supersymmetry: the untwisted sector has $\mathcal{N} = 4$ supersymmetry, while the two types of twisted sectors have $\mathcal{N} = 2$ and $\mathcal{N} = 1$ supersymmetry, if they are confined to a fixed plane\footnote{This fixed plane can be either a torus or an orbifold itself.} and a fixed point, respectively. These orbifold models have various moduli, in particular, there are always the dilaton, which controls the strength of the string coupling, and three untwisted K\"ahler moduli $T_i$ associated to the sizes of the tori. In principle, there can be also complex structure moduli $U_j$ for the three tori, if they are not fixed by the orbifold projection. For example, the `mini-landscape' models have three K\"ahler moduli $T_1, T_2, T_3$ and a complex structure modulus $U_3$ for the third complex plane.

Without matter fields, the K\"ahler potential for the moduli is given by \cite{Witten:1985xb}
\begin{equation}
  \label{Eq_ModuliTreeLevelKaehler}
  K = - \sum_{i = 1}^{h_{(1,1)}} \log (T_i + \bar{T}_i) - \sum_{j = 1}^{h_{(2,1)}} \log (U_j + \bar{U}_j) \, ,
\end{equation}
where $h_{(1,1)}$ and $h_{(2,1)}$ count the number of untwisted K\"ahler moduli $T_i$ and complex structure moduli $U_j$, respectively. There are at least three untwisted K\"ahler moduli, i.e. $h_{(1,1)} \geq 3$, the number of additional untwisted K\"ahler moduli and the number of untwisted complex structure moduli, $h_{(2,1)}$, are model-dependent. Here, we do not consider twisted moduli and focus on the three `diagonal' or `universal' K\"ahler moduli, which parametrize the volumes of the three orbifold planes.

Denoting the metric on one of the tori by $G_{i j}$, the geometric moduli $T$ and $U$ associated to this torus are given by
\begin{equation}
  \label{Eq_ModuliDefinition}
  T = \frac{1}{2} \left( \sqrt{G} + i B_{1 2} \right) \,\, , \,\, U = \frac{1}{G_{1 1}} \left( \sqrt{G} + i G_{1 2} \right) \, .
\end{equation}
$B_{i j}$ denotes the components of the 2-form in the torus. If one introduces an explicit parametrization of the metric $G_{i j}$ on this torus as follows
\begin{equation}
  \label{Eq_ModuliTorusMetric}
  G_{i j} = \begin{pmatrix} R_{1}^2 & R_1 R_2 \cos \theta_{1 2} \\ R_1 R_2 \cos \theta_{1 2} & R_{2}^2 \end{pmatrix} \, ,
\end{equation}
the moduli $T$ and $U$ are given by\footnote{Note that the radii $R_i$ are measured in string units.}
\begin{equation}
  \label{Eq_ModuliTorusExplicit}
  T = \frac{1}{2} \left( R_{1} R_{2} \sin \theta_{1 2} + i B_{1 2} \right) \,\, , \,\, U = \frac{R_2}{R_1} \sin \theta_{1 2} + i \frac{R_2}{R_1} \cos \theta_{1 2}  \, .
\end{equation}
Note that $U$ depends only on the ratio of the radii and thus determines the shape of the torus, while $T$ determines the overall size of the torus.

The Heisenberg symmetry arises at tree-level for the untwisted matter fields denoted by $\Phi_{a}^{i}$: Fixing the complex structure modulus of the $i$-th torus, the K\"ahler potential depends only on the combination \cite{Witten:1985xb}
\begin{equation}
  \label{Eq_HeisenbergInvariantCombination}
  \rho_i = T_{i} + \bar{T}_i - \sum_{a} \lvert \Phi_{a}^{i} \rvert ^2 \, .
\end{equation}
Note that it is only a symmetry of the K\"ahler potential and generically need not be preserved by the superpotential. The $\rho_i$ are related to the radii $R_i$ of the $i$-th torus in the presence of a non-trivial background for the $\Phi_{a}^i$, $\rho_i \sim R_{i}^2$, since the moduli $T_i$ have to be redefined (see e.g. \cite{LopesCardoso:1994is}).

In \cite{Ellwanger:1986yh}, the 10d origin of the Heisenberg symmetry was discussed. The symmetry appears in the limit of vanishing superpotential and gauge coupling and can be traced back to a shift of the 10d gauge fields $A_{M}^{\alpha}$ by a harmonic form $\lambda_{M}^{\alpha}$,
\begin{equation}
  \label{Eq_Heisenberg10dGaugeHarmonic}
  A_{M}^{\alpha} \rightarrow A_{M}^{\alpha} + \lambda_{M}^{\alpha} \, ,
\end{equation}
and a corresponding shift of the 2-form $B_{M N}$ by
\begin{equation}
  \label{Eq_Heisenberg10d2FormHarmonic}
  B_{M N} \rightarrow B_{M N} - \sqrt{\frac{1}{2}} \,\, A_{[ M}^{\alpha} \, \lambda_{N ]}^{\alpha} \, .
\end{equation}
Upon compactification on an orbifold these transformations induce the Heisenberg symmetry transformations on the fields $T_i$ and $\Phi_{a}^{i}$. It is expected that the Heisenberg symmetry is related to the T-duality group. It remains a task for the future to clarify this connection.

For fixed complex structure moduli $U_j$, the tree-level K\"ahler potential for the K\"ahler moduli $T_i$ and both untwisted matter fields $\Phi_{a}^{i}$ and twisted matter fields $\Psi_a$ is of the following form
\begin{equation}
  \label{Eq_TreeLevelKaehlerpotential}
  K_{0}^{\text{tree}} = - \sum_{i} \log \rho_i + \sum_{a} \left( \prod_{i} \rho_{i}^{- q_{i}^{a}} \right)  \lvert \Psi_a \rvert^2 \, ,
\end{equation}
where the exponents $q_{i}^{a}$ are determined by the corresponding twist vector (cf. Eq.~\eqref{Eq_ModularWeightsKaehler}). In general, these are rational numbers and there can be two cases: either all three $q_i$'s are non-zero or exactly one of them vanishes \cite{Dixon:1989fj}. If one expands the K\"ahler potential for the untwisted matter fields $\Phi_{a}^i$ to quadratic order, the tree-level contribution has the same form as for the twisted matter fields with $q_{j}^{a} = \delta^{i}_j$. Note that the K\"ahler potential in Eq.~\eqref{Eq_TreeLevelKaehlerpotential} is valid in the limit where $\text{Re} \, T_i$ is much larger than the matter fields $\Phi_{a}^{i}$ and $\Psi_{a}$, i.e.  $\langle \text{Re} \, T_i \rangle \gg \langle \Phi_{a}^{i} \rangle, \langle \Psi_a \rangle$.

For the dilaton, there are two different formalisms which are closely related: the string spectrum contains the antisymmetric tensor field $b_{\mu \nu}$. We can combine this field with the dilaton in a linear multiplet $L$. Alternatively, we can perform a duality transformation to implement this tensor field as an axion and describe it together with the dilaton as a chiral multiplet $S$. Both formalisms are believed to be equivalent even at the non-perturbative level \cite{Burgess:1995kp}. In the chiral multiplet formalism, the tree-level K\"ahler potential for the dilaton is given by
\begin{equation}
  \label{Eq_TreeLevelKaehlerDilatonChiral}
  K_{\text{ch}}^{\text{tree}} = - \log(S + \bar{S}) \, ,
\end{equation}
while in the linear multiplet formalism it is instead given by
\begin{equation}
  \label{Eq_TreeLevelKaehlerDilatonLinear}
  K_{\text{lin}}^{\text{tree}} = \log L \, .
\end{equation}
At tree level, the two formalisms are related by
\begin{equation}
  \label{Eq_LinearChiralTreeRelation}
  \ell = \frac{1}{s + \bar{s}} \, ,
\end{equation}
where $\ell$ and $s$ denote the lowest component of the linear and chiral multiplet, respectively. Thus, the weak-coupling limit corresponds to $\ell \rightarrow 0$ or $s \rightarrow \infty$. However, Eq.~\eqref{Eq_LinearChiralTreeRelation} is subject to both perturbative and non-perturbative corrections and we will discuss the required modifications of Eq.~\eqref{Eq_LinearChiralTreeRelation} in Sect.~\ref{Sect_NonPerturbativeCorrections}.

\subsection{Target space modular invariance}
\label{Sect_ModularInvariance}

The low-energy effective supergravity action is subject to strong constraints from target space modular invariance, which is preserved to all orders in perturbation theory. We will discuss these symmetry transformations and the restrictions it imposes on the superpotential. The latter strongly constrain the functional form of the moduli-dependent functions $a(T_i), b(T_i)$ and $c(T_i)$ in Eq.~\eqref{Eq_GeneralSuperpotential}.

The modular transformations of the K\"ahler moduli $T_i$ and complex structure moduli $U_j$ are elements of $SL(2,\mathbb{Z})$ \cite{Witten:1985xb,Shapere:1988zv}. There is one such $SL(2, \mathbb{Z})$ group for each modulus $M \in \{ T_i, U_j \}$, which acts on $M$ as
\begin{equation}
  \label{Eq_ModularTransformationModuli}
  M \rightarrow \frac{a M - i b}{i c M + d} \, , \quad a d - b c = 1 \, , \quad a, b, c, d \in \mathbb{Z} \, ,
\end{equation}
and hence
\begin{equation}
  \label{Eq_ModularTransformationLog}
  \log(M + \bar{M}) \rightarrow \log \left( \frac{M + \bar{M}}{(i c M + d)(- i c \bar{M} + d)} \right) \, .
\end{equation}
Therefore, the modular group induces a transformation of the K\"ahler potential:
\begin{equation}
  \label{Eq_ModularTransformationKaehler}
  K \rightarrow K + \sum_{i=1}^{h_{(1,1)}} \log \lvert i c_i T_i + d_i \rvert^2 + \sum_{j=1}^{h_{(2,1)}} \log \lvert i c_j U_j + d_j \rvert^2 \, ,
\end{equation}
where $h_{(1,1)}$ and $h_{(2,1)}$ again count the number of K\"ahler moduli $T_i$ and complex structure moduli $U_j$, respectively. Since the scalar potential is necessarily invariant and depends only on the combination $G = K + \log \lvert W \rvert^2$, the superpotential must also transform under modular transformations according to
\begin{equation}
  \label{Eq_ModularTransformationSuper}
  W \rightarrow \prod_{i=1}^{h_{(1,1)}} \prod_{j=1}^{h_{(2,1)}} (i c_i T_i + d_i)^{-1}(i c_j U_j + d_j)^{-1} \, W \, .
\end{equation}
Moreover, the matter fields $\Phi_{\alpha} = \{ \Phi_{a}^i , \Psi_a \}$ transform under the modular group as
\begin{equation}
  \label{Eq_ModularTransformationMatter}
  \Phi_{\alpha} \rightarrow \prod_{i=1}^{h_{(1,1)}} \prod_{j=1}^{h_{(2,1)}} (i c_i T_i + d_i)^{- q_{i}^{\alpha}}(i c_j U_j + d_j)^{- p_{j}^{\alpha}} \, \Phi_\alpha \, .
\end{equation}
The exponents $q_{i}^{\alpha}, p_{j}^{\alpha}$ in Eq.~\eqref{Eq_TreeLevelKaehlerpotential}~and~\eqref{Eq_ModularTransformationMatter} are called modular weights \cite{Dixon:1989fj,Ibanez:1992hc}. They are determined by the orbifold twist vector of the given sector $\eta_{i}(k)$, cf. Eq.~\eqref{Eq_OrbifoldTwistVector1}, as follows
\begin{subequations}
  \label{Eq_ModularWeightsKaehler}
\begin{align}
    q_{i}^{\alpha} & \equiv (1 - \eta_{i}(k)) +  N_i - \bar{N}_i & \text{for} \,\, \eta_{i}(k) \neq 0 \, , \\
    q_{i}^{\alpha} & \equiv \quad \quad N_i - \bar{N}_i & \text{for} \,\, \eta_{i}(k) = 0 \, ,
\end{align}
\end{subequations}
where the $N_i$ and $\bar{N}_i$ are integer oscillator numbers of left-moving oscillators $\widetilde{\alpha}_i$ and $\bar{\widetilde{\alpha}}_i$, respectively. Similarly, the $p_{j}^{\alpha}$ are given by
\begin{subequations}
  \label{Eq_ModularWeightsComplexStructure}
  \begin{align}
    p_{j}^{\alpha} & \equiv (1 - \eta_{j}(k)) -  N_j + \bar{N}_j & \text{for} \,\, \eta_{j}(k) \neq 0 \, , \\
    p_{j}^{\alpha} & \equiv \quad \quad - N_j + \bar{N}_j & \text{for} \,\, \eta_{j}(k) = 0 \, .
  \end{align}
\end{subequations}

For a given polynomial in the matter fields to be used in the superpotential, the correct transformation of $W$ can be ensured by appropriate powers of the Dedekind $\eta$-function multiplying this polynomial. Under modular transformations, the $\eta$-function transforms as
\begin{equation}
  \label{Eq_ModularTransformationEta}
  \eta(M) \rightarrow (i c M + d)^{1/2} \eta(M) \, ,
\end{equation}
up to a phase, where
\begin{equation}
  \label{Eq_EtaFunction}
  \eta(M) = \exp \left(- \frac{\pi M}{12} \right) \prod_{n=1}^{\infty} \left( 1 - e^{- 2 \pi n M} \right) \, .
\end{equation}
Thus, a generic term in the superpotential has the following structure
\begin{equation}
  \label{Eq_GenericTermSuperpotential}
  W \supset \prod_{\alpha} \Phi_{\alpha}^{n_{\alpha}} \prod_{i = 1}^{h_{(1,1)}} \eta(T_i)^{2 \sigma_i}  \prod_{j = 1}^{h_{(2,1)}} \eta(U_j)^{2 \widetilde{\sigma}_j} \, ,
\end{equation}
where $\sigma_i = - 1 + \sum_{\alpha}  n_{\alpha} q_{i}^{\alpha}$ and $\widetilde{\sigma}_j = - 1 + \sum_{\alpha}  n_{\alpha} p_{j}^{\alpha} $. The index $\alpha$ runs over both untwisted and twisted matter fields. For $\text{Re} \, T_i \gtrsim 1$, we can approximate $\eta(T_i)$ by $\exp( - \frac{\pi \, T_i}{12} )$ and hence if a term in the superpotential has a moduli-dependence, it is generically of the form $\sim e^{- c \, T_i}$ at large radius (i.e. for large $\text{Re} \, T_i$) for some constant $c$. If a term contains an explicit factor of $e^{- c T_i}$, it is interpreted as being generated by non-perturbative effects: The strings have so stretch over the $i$-th torus to reach each other, which leads to a suppression by the volume. In principle, there are terms in $W$ which do not depend on the moduli (up to a modular invariant function which we do not consider here). For instance, three untwisted fields associated to three different planes or three twisted fields living at the same fixed point will not have any moduli dependence.\footnote{Up to a modular invariant function, which we do not consider here.}

As is well known, the superpotential for the matter fields starts at cubic order in the fields after the heavy string states have been integrated out. This is why we introduced the field $\Sigma$ in Eq.~\eqref{Eq_GeneralSuperpotential} to generate the F-term of $X$ by acquiring an expectation value. In general, $\Sigma$ will be some product of fields, which we collectively denoted by $\Sigma^2$ in Eq.~\eqref{Eq_GeneralSuperpotential}. Similarly, also the functions $a(T_i), b(T_i)$ and $c(T_i)$ might depend on expectation values of some matter fields, which will affect their moduli dependence. We discuss this issue in more detail in Sect.~\ref{Sect_RealizationFieldContent}.

\subsection{Gauge kinetic function and Green-Schwarz counterterm}
\label{Sect_LoopCorrectionsGaugeKinetic}

Here, we review the gauge kinetic function and its string one-loop corrections. The modular transformations are anomalous at the string one-loop level and this anomaly is cancelled by the Green-Schwarz mechanism and threshold corrections from massive string modes, which in turn modifies the effective action. In particular, in the chiral multiplet formalism the dilaton $S$ will generically mix with the K\"ahler moduli $T_i$ and the complex structure moduli $U_j$. This mixing makes finding flat directions suitable for inflation more complicated.

In the chiral multiplet formalism, the gauge couplings are determined by $g_{a}^{-2} = \text{Re} \, f_a$ with the gauge kinetic function $f_a$. At string one-loop level, $f_a$ is given by \cite{Derendinger:1991hq,Derendinger:1991kr,Lust:1991yi,Kaplunovsky:1995jw}
\begin{equation}
  \label{Eq_LoopCorrectionsGauge}
  f_{a}(S,T) = k_{a} S + \sum_{i=1}^{h_{(1,1)}} \left( \alpha_{a}^i - k_a \delta_{GS}^i \right) \log ( \eta(T_i) )^2  + \sum_{j=1}^{h_{(2,1)}} \left( \alpha_{a}^j - k_{a} \delta_{GS}^j \right) \log ( \eta(U_j
) )^2 \, ,
\end{equation}
where $a$ labels the different gauge groups, $k_a$ is the Kac-Moody level of the group (typically $k_a = 1$), and the model-dependent constants $\alpha_{a}^i$ are defined as
\begin{equation}
  \label{Eq_LoopCorrectionsCoeff1}
  \alpha_{a}^i \equiv \ell(\text{adj}) - \sum_{\text{rep}_A} \ell_{a}(\text{rep}_A) ( 1 + 2 q_{i}^A ) \, .
\end{equation}
Here, $\ell(\text{adj})$ and $\ell_{a}(\text{rep}_A)$ are the Dynkin indices of the adjoint and matter field representations of the corresponding gauge group $\mathcal{G}_a$, respectively.\footnote{The Dynkin indices are determined from the normalization condition $Tr(T_i T_j) = \ell_{a}(\text{rep}) \delta_{i j}$ of the generators $T_i$ in the given representation.} The coefficients $\delta_{GS}^i$ are given by \cite{Kaplunovsky:1995jw}
\begin{equation}
  \label{Eq_LoopCorrectionsCoeff2}
  \alpha_{a}^i - k_a \delta_{GS}^i = \frac{b_{a}^{i, \, \mathcal{N}=2}}{\lvert D \rvert / \lvert D_i \rvert} \, ,
\end{equation}
where $b_{a}^{i, \, \mathcal{N}=2}$ is a beta function coefficient of the gauge group $G_a$ for the $i$-th torus. These coefficients are non-zero only if there is some twisted sector with $\mathcal{N} = 2$ supersymmetry and if this twisted sector does not rotate the $i$-th torus. The factors $\lvert D \rvert$ and $\lvert D_i \rvert$ are the degree of the twist group $D$ and the little group $D_i$, which leaves the $i$-th unrotated, respectively. For example, the mini-landscape models have $D = \mathbb{Z}_{6-II}$ and $\lvert D \rvert = 6$, $\lvert D_2 \rvert = 2$ and $\lvert D_3 \rvert = 3$ since the little groups under which the second and third torus are fixed are $\mathbb{Z}_2$ and $\mathbb{Z}_3$, respectively. The first torus is rotated in all twisted sectors.

The $\delta_{GS}^i$ terms are introduced to cancel a sigma-model and K\"ahler anomaly of the modular group. This anomaly induces a non-trivial modular transformation of the dilaton in the chiral formalism:
\begin{equation}
  \label{Eq_DilatonAnomalousTransformation}
  S \rightarrow S + \sum_{i = 1}^{h_{(1,1)}} \delta_{GS}^i \log(i c_i T_i + d_i) + \sum_{j = 1}^{h_{(2,1)}} \delta_{GS}^j \log(i c_j U_j + d_j) \, .
\end{equation}
This anomaly is cancelled (partially) by the so-called Green-Schwarz counterterm, which modifies the K\"ahler potential at string one-loop level. Neglecting the matter fields, the modified 
K\"ahler potential in the chiral multiplet formalism is given by
\begin{equation}
  \label{Eq_GreenSchwarzKaehler1}
  K = - \log Y - \sum_{i = 1}^{h_{(1,1)}} \log (T_i + \bar{T}_i) - \sum_{j=1}^{h_{(2,1)}} \log ( U_j + \bar{U}_j) \, ,
\end{equation}
where
\begin{equation}
  \label{Eq_GreenSchwarzKaehler2}
  Y = S + \bar{S} - \sum_{i = 1}^{h_{(1,1)}} \delta_{GS}^i \log (T_i + \bar{T}_i) - \sum_{j=1}^{h_{(2,1)}} \delta_{GS}^j \log ( U_j + \bar{U}_j) \, .
\end{equation}
Thus, generically the dilaton mixes with the K\"ahler and complex structure moduli. This makes computations in the chiral formalism somewhat more complicated, in particular, the diagonalization of the kinetic terms, which is necessary when discussing flat directions. In the linear multiplet formalism, however, the dilaton is inert under modular transformations.\footnote{One can make a field redefinition of the dilaton in the chiral formalism in order to keep it inert under modular transformations, cf. e.g. \cite{Derendinger:1991hq}.} In this formalism, the Green-Schwarz counterterm is implemented differently \cite{Derendinger:1991hq,LopesCardoso:1991zt}. Neglecting the complex structure moduli, it is determined by the quantity
\begin{equation}
  \label{Eq_GreenSchwarzLinear1}
  V_{GS} = - \sum_{i} \delta_{GS}^i \log ( T_i + \bar{T}_i ) \, .
\end{equation}
Here, we will assume that $V_{GS}$ preserves the Heisenberg symmetry, i.e. that it is actually given by the tree-level K\"ahler potential \cite{Gaillard:1998xx}:
\begin{equation}
  \label{Eq_GreenSchwarzLinear2}
  V_{GS} =  - \sum_{i} \delta_{GS}^i \log \rho_i + \sum_{a} p_{a} \lvert \Psi_a \rvert^2 \left( \prod_{i} \rho_{i}^{- q_{i}^{a}} \right) \, ,
\end{equation}
with the unknown contribution of the twisted matter fields $\Psi_a$ to the Green-Schwarz term parametrized by the coefficients $p_a$. Upon including this term, the effective K\"ahler metric for the fields is modified to
\begin{equation}
  \label{Eq_GreenSchwarzEffectiveKaehler}
  K_{m \bar{n}}^{\, \text{eff}} = K_{m \bar{n}} + \ell V^{GS}_{m \bar{n}} \, .
\end{equation}
In general, the Green-Schwarz mechanism will not cancel the complete modular anomaly completely. The remaining part of the anomaly is cancelled by threshold corrections from massive string modes \cite{Kaplunovsky:1995jw}. These threshold corrections are moduli-dependent since the masses of e.g. the Kaluza-Klein and winding states depend on the radii.

\subsection{Loop corrections to the matter K\"ahler potentials}
\label{Sect_LoopCorrectionsKaehler}

The stabilization of the moduli during inflation, as explained in Sect.~\ref{Sect_GeneralModuli}, relies crucially on the function $d(\rho_3)$. In this section, we propose a functional form of $d(\rho_3)$, which we expect to arise in heterotic orbifold compactifications. First, we consider known results for the string one-loop corrections to the K\"ahler metric of untwisted matter fields. Based on these results, we suggest a generalization in the presence of background values for matter fields and argue why we expect the Heisenberg symmetry to be preserved in the large radius limit, up to exponentially suppressed terms.

The result of \cite{Antoniadis:1992pm} for the string one-loop corrections to the K\"ahler metric of untwisted matter fields in $\mathcal{N} = 2$ orbifolds has the following form:
\begin{equation}
  \label{Eq_LoopKaehler1}
  K_{n \bar{n}}^{\, \text{eff}} = K_{n \bar{n}}^{\, \text{tree}}  + \ell K_{n \bar{n}}^{\, \text{1-loop}} \, ,
\end{equation}
with
\begin{equation}
  \label{Eq_LoopKaehler2}
  K_{n \bar{n}}^{\text{1-loop}} = K_{n \bar{n}}^{\, \text{tree}}  \left( \gamma + \beta \, Y(T, \bar{T}) \right) \, ,
\end{equation}
where
\begin{equation}
  \label{Eq_LoopKaehler3}
  Y(T, \bar{T}) = \log \left[ \lvert \eta(T) \rvert^4 (T + \bar{T}) \right] \, ,
\end{equation}
and $\ell$ is the loop counting parameter, the lowest component of $L$ or $\sim (S + \bar{S})^{-1}$. $\beta$ is related to the $\mathcal{N} = 2$ beta function coefficient of the torus associated with $T$ and the gauge group which acts non-trivially on the matter field under consideration. The moduli-independent constant $\gamma$ is the effect of the $\mathcal{N} = 1$ subsectors. The moduli-dependence of $Y(T, \bar{T})$ in Eq.~\eqref{Eq_LoopKaehler2} originates in loops involving massive string states, namely Kaluza-Klein and winding modes, whose masses depend on the moduli, especially on the radius. Note that the moduli-dependent correction $Y(T, \bar{T})$ arises only from $\mathcal{N} = 2$ sectors, which leave the plane associated to the matter field unrotated and therefore only depends on the moduli of that plane.

For $\text{Re} \, T \gtrsim 1$, we can approximate $Y(T, \bar{T})$, Eq.~\eqref{Eq_LoopKaehler3}, by
\begin{equation}
  \label{Eq_LoopKaehlerModuliDependence1}
  Y(T, \bar{T}) \approx \log(T + \bar{T}) - \frac{\pi}{6} (T + \bar{T}) + \mathcal{O}(e^{- 2 \pi T}) + c.c.  \, ,
\end{equation}
as can be easily seen from Eq.~\eqref{Eq_EtaFunction}. The dependence on $\text{Im} \, T$ is only through the additional terms $\sim e^{- 2 \pi T}$, which are exponentially suppressed for large $\text{Re} \, T$, i.e. for a large compactification radius. In other words, the continuous shift symmetry $T \rightarrow T + i \alpha$  (which is broken to a discrete one by worldsheet instantons \cite{Dine:1986zy,Ferrara:1991uz}) survives as an approximate symmetry in the large radius limit.

Based on these results, we propose a generalization involving background values for untwisted matter fields as described below. It remains a task for the future to check our proposal by calculating the relevant string amplitudes.

As a working hypothesis, we consider the case where we simply replace $T + \bar{T}$ in the large radius limit of $Y(T, \bar{T})$, Eq.~\eqref{Eq_LoopKaehlerModuliDependence1}, by $\rho \equiv T + \bar{T} - \sum_{a} \lvert \Phi_a \rvert^2$, i.e.
\begin{equation}
  \label{Eq_LoopKaehlerModuliDependence2}
  Y(T, \bar{T}, \Phi_a, \bar{\Phi}_a) = \log \rho - \frac{\pi}{6} \rho + \mathcal{O}(e^{- \pi \rho}) \, ,
\end{equation}
and only the exponentially suppressed terms break the Heisenberg symmetry. In the following, we parametrize this breaking by a term $\lambda \sum_{a} \lvert \Phi_a \rvert^2$, with $\lambda \sim e^{- \pi \rho}$ exponentially small for large radius (i.e. large $\rho$). The coefficient $\lambda$ has to be computed directly from string amplitudes, and in general will depend on $T$. The assumption that the Heisenberg symmetry is broken only by exponentially suppressed terms in the large radius limit is based on the observation that this happens for the continuous shift symmetry $T \rightarrow T + i \alpha$, which is part of the Heisenberg symmetry group, cf. Eqs.~\eqref{Eq_GeneralHeisenberg1} and \eqref{Eq_GeneralHeisenberg2}.

We now apply this assumption to the setup considered in Sect.~\ref{Sect_GeneralModel}, that is we consider
\begin{equation}
  \label{Eq_LoopKaehlerTwisted}
  K_{X \bar{X}} =  \left( \prod_{j = 1}^{2} (T_{j} + \bar{T}_{j})^{- q_{j}} \right) \left[ 1 + \ell \gamma + \ell \beta_3 \left( \log \rho_3 - \frac{\pi}{6} \rho_3 + \lambda \sum_a \lvert \Phi_a \rvert^2 \right) \right] \, ,
\end{equation}
which is of the form Eq.~\eqref{Eq_GeneralXkinetic3} with $d(\rho_3) = Y(T_3, \bar{T_3}, \Phi_a, \bar{\Phi}_a)$ given by Eq.~\eqref{Eq_LoopKaehlerModuliDependence2}.

\subsection{Non-perturbative corrections}
\label{Sect_NonPerturbativeCorrections}

So far, we have described the structure of the effective supergravity theory at tree-level and introduced perturbative corrections. Now we will introduce also non-perturbative corrections. These are an important ingredient for successful moduli stabilization in string theory and in heterotic models they are crucial in order to stabilize the dilaton. The stabilization scheme which we will employ here is known as  K\"ahler stabilization \cite{Binetruy:1996xja}. It relies on non-perturbative corrections to the superpotential from gaugino condensation in combination with non-perturbative corrections to the K\"ahler potential. These non-perturbative corrections can be either of a field-theoretic \cite{Banks:1994sg} or a stringy origin \cite{Shenker:1990uf} (see also \cite{Polchinski:1994fq}). Field-theory instantons scale like $e^{- 1 / g^2}$, while string theory instanton effects scale like $e^{- 1 / g}$, where $g$ is the coupling constant.

In the following, we first briefly comment on non-perturbative effects in the chiral multiplet formalism and then we turn to their description in terms of the linear multiplet. The discussion here follows \cite{Binetruy:1996xja}.

\subsubsection*{The chiral multiplet formalism}

In the chiral multiplet formalism, one has non-perturbative corrections to both the K\"ahler and the superpotential,
\begin{equation}
  \label{Eq_ChiralNonPerturbative}
  \begin{split}
  K & = K_{\text{tree}} + K_{\text{pert}} + K_{\text{np}} \, , \\
  W & = W_{\text{tree}} + W_{\text{np}} \, .
  \end{split}
\end{equation}
The non-perturbative superpotential is due to the presence of a gaugino condensate and thus one has \cite{Lust:1991yi,Font:1990nt}
\begin{equation}
  \label{Eq_ChiralGauginoSuper}
  W_{\text{np}} = A e^{- b S} \prod_{i=1}^{3} \eta(T_i)^{-2} \, ,
\end{equation}
where $b$ is related to the beta-function coefficient of the condensing gauge group and the $\eta$-functions are introduced to ensure covariance of the superpotential under modular transformations. The non-perturbative corrections to the K\"ahler potential are typically parametrized in terms of $\text{Re} \, S$, cf. e.g. \cite{Casas:1996zi} for some examples.

The kinetic mixing between the dilaton and the K\"ahler moduli, cf. Eq.~\eqref{Eq_GreenSchwarzKaehler1}, however makes finding flat directions more complicated in the chiral multiplet formalism. Thus, we will focus on the linear multiplet formalism and since the two formalisms are believed to be equivalent there should be no physical difference.

\subsubsection*{The linear multiplet formalism}

In the linear multiplet formalism, the superpotential is independent of the dilaton since it is not a chiral superfield. The non-perturbative corrections to the K\"ahler potential are parametrized by a function $g(L)$ as
\begin{equation}
  \label{Eq_LinearNonPerturbativeKaehler}
  K = \log L + g(L) + \dots \, ,
\end{equation}
where the dots denote the terms involving the other moduli and matter fields. The gauge coupling constant (at the string scale) also receives non-perturbative corrections given by another function $f(\ell)$:
\begin{equation}
  \label{Eq_LinearGaugeCoupling}
  g^{2} = \frac{2 \ell}{1 + f(\ell)} \, .
\end{equation}
The relation between the linear and the chiral multiplet formalism gets modified by both perturbative and non-perturbative effects \cite{Binetruy:1996xja}:
\begin{equation}
  \label{Eq_LinearChiralRelation}
  \frac{\ell}{1 + f(\ell)} = \frac{1}{s + \bar{s} + V_{GS}} \, ,
\end{equation}
where $V_{GS}$ is given by Eq.~\eqref{Eq_GreenSchwarzLinear2}.

The two functions $g(\ell)$ and $f(\ell)$ in Eqs.~\eqref{Eq_LinearNonPerturbativeKaehler} and \eqref{Eq_LinearGaugeCoupling} are related by
\begin{equation}
  \label{Eq_LinearFunctionRelation}
  \ell g^\prime = f - \ell f^\prime \, , \quad f(\ell = 0) = g(\ell =0) = 0 \, ,
\end{equation}
where the prime denotes a derivative with respect to $\ell$. The differential equation and boundary condition ensure canonical normalization of the Einstein term and the correct behaviour in the weak-coupling limit $\ell \rightarrow 0$, respectively. Following \cite{Binetruy:1996xja}, we parametrize $f(\ell)$ as
\begin{equation}
  \label{Eq_LinearNonPerturbativeFunction}
  f(\ell) = B \left( 1 + A \frac{1}{\sqrt{a \ell}} \right) e^{- 1/ \sqrt{a \ell}} \, .
\end{equation}

During inflation, the gaugino condensate is expected to be negligible and hence the effective scalar potential for vanishing D-terms is given by \cite{Gaillard:1998xx}
\begin{equation}
  \label{Eq_LinearEffectivePotential}
  V =  e^{K} \left( (\ell g^{\prime}(\ell) + 1) \lvert W \rvert^2 - 3 \lvert W \rvert^2 + \sum_{m \bar{n}} \left(K_{m \bar{n}}^{\, \text{eff}} \right)^{-1} F_{m} \bar{F}_{\bar{n}} \right) \, ,
\end{equation}
where the indices $m, n$ run over the scalar components of the K\"ahler moduli $T_i$, the untwisted matter fields $\Phi_{a}^{i}$ and the twisted matter fields $\Psi_a$ and with $F_m$ given by
\begin{equation}
  \label{Eq_LinearEffectiveFterm}
  F_{m} = W_m + K_{m} W \, .
\end{equation}
The effective K\"ahler metric in the last term of Eq.~\eqref{Eq_LinearEffectivePotential} is given by
\begin{equation}
  \label{Eq_LinearEffectiveKaehler}
  K_{m \bar{n}}^{\text{eff}} = K_{m \bar{n}}^{\text{tree}} + \ell K_{m \bar{n}}^{\text{1-loop}} \, ,
\end{equation}
while the $K$ to be used in Eqs.~\eqref{Eq_LinearEffectivePotential} and \eqref{Eq_LinearEffectiveFterm} is given by
\begin{equation}
  \label{Eq_LinearTreeKaehler}
  K = \log(\ell) + g(\ell) - \sum_{i} \log \rho_i + \sum_{a} \left( \prod_{i} \rho_{i}^{- q_{i}^{a}} \right)  \lvert \psi_a \rvert^2 \, ,
\end{equation}
which is obtained by replacing all superfields with their scalar components (denoted by lower case letters) and dropping the perturbative corrections. Note that in Eq.~\eqref{Eq_LinearTreeKaehler} we use $\rho_i = t_i + \bar{t}_i - \sum_{a} \lvert \phi_{a}^{i} \rvert^2$.

\subsection{Anomalous $U(1)_A$ and generating expectation values}
\label{Sect_AnomalousU1andVEVs}

In this section, we review how to generate expectation values for matter fields via D-terms of an anomalous $U(1)_A$ and F-terms. For more details and examples in the present context of inflation model building, see e.g. \cite{Gaillard:1998xx,Kain:2006nx}. Recall that we introduced the field $\Sigma$ in Eq.~\eqref{Eq_GeneralSuperpotential}, which has to acquire an expectation value in order to generate an F-term for $X$. This is necessary since the string theory superpotential starts at cubic order in the matter fields and thus no linear terms are present unless some fields acquire expectation values. These are collectively represented by $\Sigma$.

\subsection*{D-term expectation values}

In many orbifold models there exists an anomalous $U(1)_A$. The anomaly is cancelled via a Green-Schwarz counterterm, which gives rise to a Fayet-Iliopoulos contribution to the D-term $D_A$. Thus, we have a contribution to the scalar potential from the D-term, 
\begin{equation}
  \label{Sect_AnomalousDtermpotential1}
  V_{D} = \frac{g^2}{2}  \left( \sum_{\alpha} q_{A}^{\alpha} K_{\alpha} \phi_{\alpha} + \xi_A \right)^2 \, ,
\end{equation}
where the index $\alpha$ runs over both twisted and untwisted matter fields, $q_{A}^{\alpha}$ denotes the charge under the anomalous $U(1)_A$ (not to be confused with the modular weights $q_{i}^{\alpha}$), the gauge coupling $g^2$ is given in Eq.~\eqref{Eq_LinearGaugeCoupling}, $\phi_{\alpha}$ denotes the scalar component of $\Phi_\alpha \in \{ \Phi_{a}^{i}, \Psi_a \}$ and the Fayet-Iliopoulos D-term $\xi_A$ (in the linear multiplet formalism) is given by
\begin{equation}
  \label{Sect_AnomalousFIterm}
  \xi_A = \frac{2 \ell \, \text{Tr} \, Q_{A}}{192 \pi^2} \, ,
\end{equation}
with $Q_{A}$ the generator of the anomalous $U(1)_A$. Using the K\"ahler potential of Eq.~\eqref{Eq_TreeLevelKaehlerpotential}, the D-term potential in Eq.~\eqref{Sect_AnomalousDtermpotential1} becomes
\begin{equation}
  \label{Sect_AnomalousDtermpotential2}
  V_{D} = \frac{1}{2} \, g^2 \left[  \sum_{\alpha} \left( \prod_{i} \rho_{i}^{- q_{i}^{\alpha}} \right) q_{A}^{\alpha} \lvert \phi_{\alpha} \rvert^2  + \xi_A \right]^2 \, .
\end{equation}
Cancellation of the D-term requires some matter fields to pick up non-zero expectation values of the form
\begin{equation}
  \label{Eq_AnomalousDtermVEV}
  \frac{\lvert \langle \phi_{\alpha} \rangle \rvert^2}{q_{A}^{\alpha}} = const \cdot \ell \cdot \left( \prod_{i} \rho_{i}^{ q_{i}^{\alpha}} \right)  \, .
\end{equation}

\subsection*{F-term expectation values}

Via the superpotential, such D-term expectation values can induce other expectation values. To illustrate this, let us review the example of \cite{Kain:2006nx}. Consider the following modular invariant expression of the three fields $\chi, \phi, \phi^{\prime}$:
\begin{equation}
  \label{Eq_FtermVEV1}
  \Gamma = \chi \phi \phi^{\prime} \prod_{i} \eta(T_i)^{\, 2 \sum_{\beta} q_{i}^{\beta}} \, , \, \beta = \chi, \phi, \phi^{\prime} \, ,
\end{equation}
and assume that $\phi$ and $\phi^{\prime}$ acquire non-zero expectation values, e.g through the cancellation of a D-term as described above. Using this expression, we can build a superpotential contribution of the form
\begin{equation}
  \label{Eq_FtermVEV2}
  W(\Gamma) = \left( \psi \phi \phi^{\prime} \prod_{i} \eta(T_i)^{-2 ( 1 - \sum_{\gamma} q_{i}^{\gamma})} \right) \sum_{n=0} c_{n} \Gamma^n \, , \, \gamma = \psi, \phi, \phi^{\prime} \, ,
\end{equation}
with some constants $c_n$, which is allowed by all the symmetries, if the products $\psi \phi \phi^{\prime}$ and $\chi \phi \phi^{\prime}$ are gauge invariant. The F-term equations can be satisfied if $\langle \psi \rangle = 0$ and
\begin{equation}
  \label{Eq_FtermVEV3}
  \sum_{n=0} c_n \Gamma^n = 0 \, .
\end{equation}
If $c_0, c_1 \neq 0$, the only solution to this equation is $\Gamma = const$ and hence
\begin{equation}
  \label{Eq_FtermVEV4}
  \lvert \langle \chi \rangle \rvert^2 = const \cdot  \Big\lvert \langle \phi \phi^{\prime} \prod_{i} \eta(T_i)^{\, 2 \sum_{\beta} q_{i}^{\beta}} \rangle \Big\rvert^{-2} \, , \, \beta = \chi, \phi, \phi^{\prime} \, .
\end{equation}
Note that if $\langle \phi \rangle$ and $\langle \phi^{\prime} \rangle$ are induced by the D-term cancellation as above, we see from Eqs.~\eqref{Eq_FtermVEV4} and \eqref{Eq_AnomalousDtermVEV} that $\lvert \langle \chi \rangle \rvert^2 \propto \ell^{-2}$. Also note that in principle $\langle \chi \rangle$ can involve $\eta(T_i)$ to some power.

\section{Realisation in heterotic orbifolds}
\label{Sect_Realization}

After reviewing the basic ingredients of the effective field theory of heterotic orbifolds above in Sect.~\ref{Sect_EffectiveAction}, we now outline how the general scenario of Sect.~\ref{Sect_GeneralModel} may be realised within such compactifications. We begin by identifying the field content, which needs to be extended to include the dilaton. We also comment on the form of the superpotential in Sect.~\ref{Sect_RealizationFieldContent}. Next, we describe how the moduli can be stabilized within this setup in Sect.~\ref{Sect_RealizationModuliStabilization}. In Sect.~\ref{Sect_RealizationDflatness}, we briefly comment on constraints from imposing D-flatness. Finally, we discuss possibilities for generating a slope and the hybrid mechanism for ending inflation in Sect.~\ref{Sect_RealizationSlope}.

\subsection{The field content and the superpotential}
\label{Sect_RealizationFieldContent}

We first identify the field content of the general class of models of Sect.~\ref{Sect_Generalization} within heterotic orbifold compactifications. Then, we discuss the requirements on the moduli dependence of the scalar potential, which is affected by the constraints on the superpotential from modular invariance and the moduli-dependence of expectation values of matter fields (cf. Sects.~\ref{Sect_ModularInvariance} and \ref{Sect_AnomalousU1andVEVs}, respectively).

The identification of the field content of Sect.~\ref{Sect_Generalization} is straightforward: The moduli $T_i$ are of course identified with the three untwisted K\"ahler moduli present in any heterotic orbifold, which determine the radii of the three tori. $f(\Phi_a)$ is a product of untwisted matter fields $\Phi_a$, associated w.l.o.g. to the third torus with modulus $T_3$, which forms a D-flat direction, e.g. $f(\Phi_a) = \Phi^+ \Phi^-$. We neglect the complex structure moduli $U_j$ here, assuming that they are either fixed by the orbifold projection or in a similar way to the $T_i$. We also neglect any twisted moduli or additional `off-diagonal' K\"ahler moduli.

We take $X$ to be a twisted matter field in a twisted sector with $\mathcal{N} = 2$ supersymmetry. This is necessary to have a non-trivial moduli dependence in its K\"ahler metric, which is parametrized by $d(\rho_3)$ in Eq.~\eqref{Eq_GeneralXkinetic3}. We expect this moduli dependence to arise from string threshold corrections, cf. Sect.~\ref{Sect_LoopCorrectionsKaehler}, and we assume here that it preserves the Heisenberg symmetry up to terms which are exponentially small in the large radius limit. The function $d(\rho_3)$ is given by Eq.~\eqref{Eq_LoopKaehlerTwisted} and is of a similar form as the simple example discussed in \ref{Sect_GeneralModuli}: in addition to the part linear in $\rho_3$ it contains a logarithmic contribution. Moreover, all these corrections are of course proportional to the dilaton $\ell$ since this is the string loop counting parameter. In order to receive moduli-dependent corrections to its K\"ahler metric, $X$ must be charged under (part of) the gauge group of the $\mathcal{N} = 2$ subsector. We will for simplicity assume that the inflaton (and the waterfall) fields are neutral with respect to this gauge group. Recall also that the inflationary setup of Sect.~\ref{Sect_GeneralModel} requires a negative quartic $\lvert X \rvert^4$ term in the K\"ahler potential and we assume that this term exists. However, the K\"ahler potential of twisted matter fields is only known to quadratic order so far and our two assumptions on the K\"ahler potential terms involving $X$ need to be checked in the future.

The superpotential starts at cubic order in the matter fields and thus the F-term of $X$ has to arise from non-vanishing expectation values for some other fields, which were collectively denoted by $\Sigma$ in  Eq.~\eqref{Eq_GeneralSuperpotential}. As reviewed in Sect.~\ref{Sect_AnomalousU1andVEVs}, these expectation values are generically moduli-dependent and thus modify the dependence of the effective scalar potential on the moduli fields. Recall that an important property of the setup of Sect.~\ref{Sect_GeneralModel} is that during inflation only $W_X \neq 0$ and $W \simeq W_m \simeq 0$ for all $m \neq X$. This property is well-suited for realisations of inflation in heterotic orbifolds \cite{Stewart:1994ts,Gaillard:1998xx}. We therefore require that a superpotential with the structure of Eq.~\eqref{Eq_GeneralSuperpotential} is present. It remains a task for the future to find explicit compactifications where this structure is realised and in particular whether this is possible in phenomenologically interesting setups such as the mini-landscape models \cite{Buchmuller:2005jr}. However, this seems possible since we can e.g. allow for terms of the form $\Phi_{a} \Psi \Psi^{\prime}$ in the superpotential if $\langle \Psi \rangle \simeq \langle \Psi^{\prime} \rangle \simeq 0$ during inflation.

The functions $a(T_i), b(T_i)$ and $c(T_i)$ in Eq.~\eqref{Eq_GeneralSuperpotential} are constrained by modular invariance: They are given by appropriate powers of the $\eta$-function $\eta(T_i)$, cf. Eq.~\eqref{Eq_GenericTermSuperpotential}. For example, we must have (for $\text{Re} \, T_3 \gtrsim 1$)
\begin{equation}
  \label{Eq_RealizationcT3dependence}
  c(T_i) \propto \eta(T_3)^{c_3} \sim e^{- \frac{\pi \, c_{3}}{12} T_3} \, ,
\end{equation}
since $f(\Phi_a)$ involves at least two untwisted fields from the same sector. Note that in general it will also depend on $T_1$ and $T_2$. During inflation, $\langle H^\pm \rangle  = 0$ and therefore the moduli dependence of $b(T_i)$ and $c(T_i)$ enters the effective scalar potential only through loops involving the waterfall fields.

There is a complication compared to the scenario of Sect.~\ref{Sect_GeneralModel}, namely that the functions $a(T_i), b(T_i)$ and $c(T_i)$ in principle might involve the expectation values of some matter fields, which can alter their moduli dependence. In particular, $\langle \Sigma \rangle$ directly affects the moduli dependence of the scalar potential and thus the stabilization of the moduli because it can depend on the moduli, as explained in Sect.~\ref{Sect_AnomalousU1andVEVs}. There are two requirements which the scalar potential has to fulfill during inflation: it must be sufficiently flat at tree-level along the inflationary trajectory and it must lead to successful moduli stabilization. As we will see below in Sect.~\ref{Sect_RealizationModuliStabilization}, the latter requirement imposes a constraint on the functional form of $\lvert a(T_i) \rvert^2 \lvert \langle \Sigma \rangle \rvert^4$ (which encodes the moduli dependence of the F-term of $X$) to be independent of $\eta(T_3)$ and to depend only on inverse powers of both $\eta(T_1)$ and $\eta(T_2)$. This can be translated into a requirement on the modular weights of certain fields. The first requirement of a sufficiently flat tree-level potential for the inflaton will be discussed in Sect.~\ref{Sect_RealizationSlope}.

\subsection{Stabilization of the dilaton and the K\"ahler moduli}
\label{Sect_RealizationModuliStabilization}

It is important to stabilize all of the moduli during inflation (preferably with masses $m \sim H_{inf}$), since none of them is considered to be the inflaton. The stabilization works essentially the same way as in the phenomenological approach in Sect.~\ref{Sect_GeneralModuli}, even though the dilaton complicates the situation significantly. We will now discuss the moduli dependence of the scalar potential and argue that all moduli can be stabilized. In particular, we show that one can stabilize $\rho_3$, as defined in Eq.~\eqref{Eq_HeisenbergInvariantCombination}, at a large value.

\begin{figure}[ht]
\begin{center}
\centerline{
\includegraphics[scale=0.67]{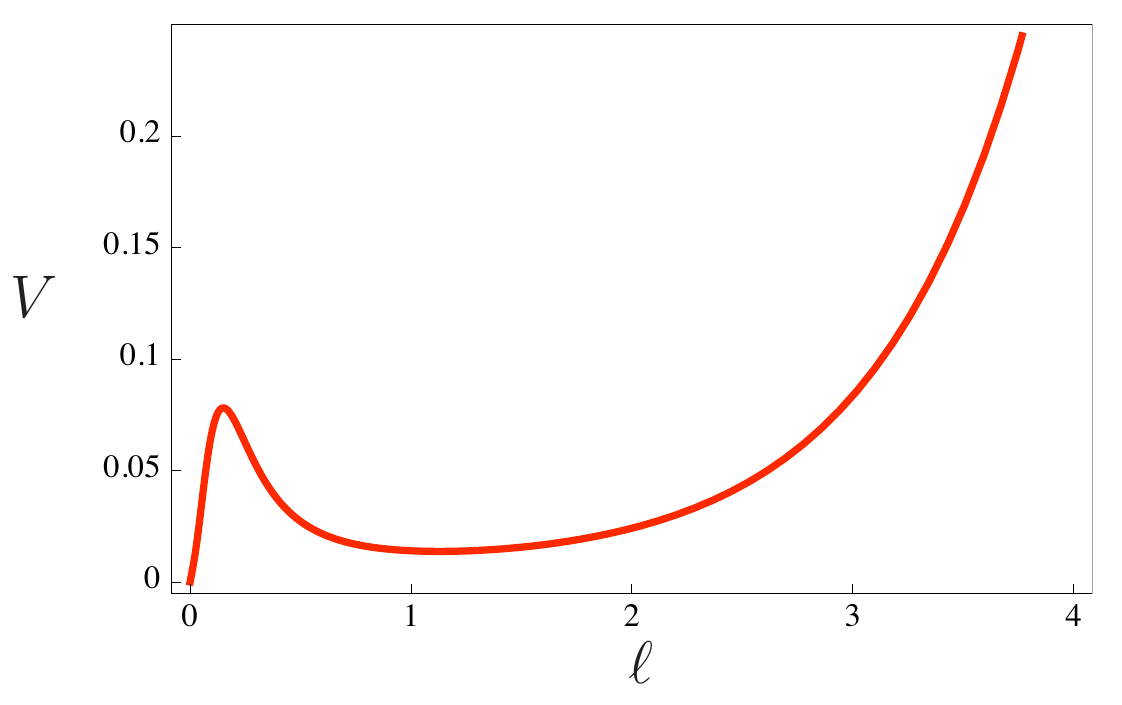}
\hspace*{0.4cm}
\includegraphics[scale=0.67]{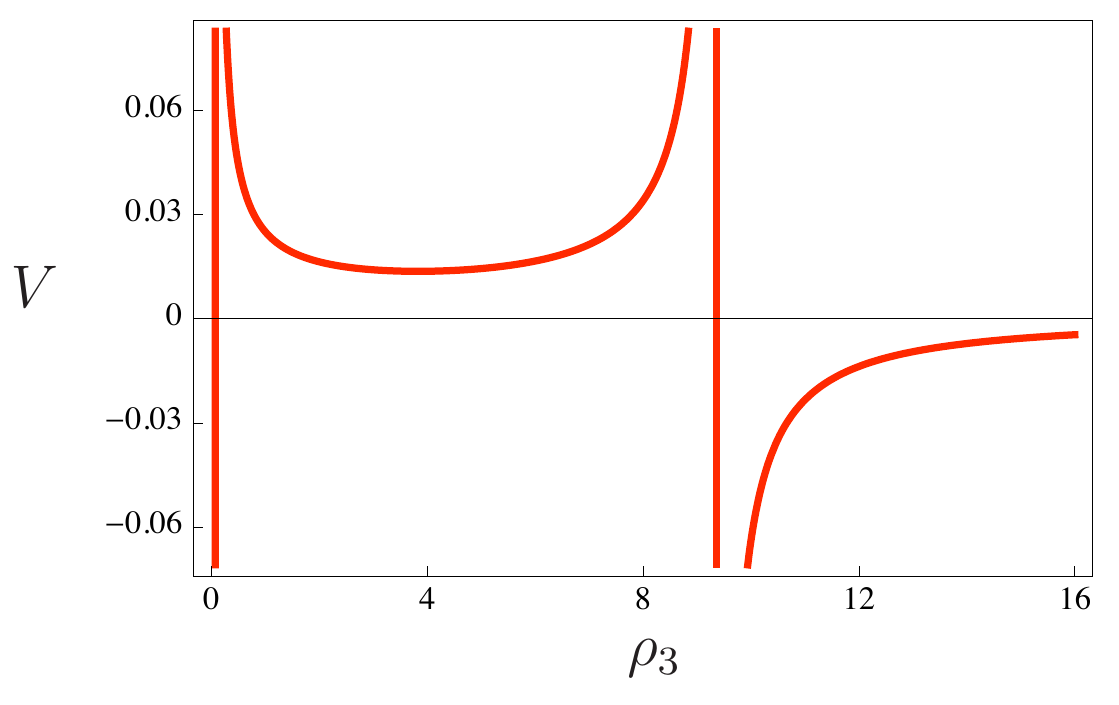}
}
\caption{Dependence of the potential Eq.~\eqref{Eq_RealizationPotentialDilatonrho3} on $\ell$ with $\rho_3$ at its minimum, and vice versa. For the example values $n = 1$, $q = - \frac{1}{2}$, $\gamma = \frac{10}{8 \pi^2} \approx 0.13$, $\beta_3 = \frac{30}{8 \pi^2} \approx 0.38$, $A = - 0.7$, $B = 20$ and $a = 1$, there is a minimum at $\langle \ell \rangle \approx 1.13 $ with $g^2 \approx 0.62$ and $\langle \rho_3 \rangle \approx 3.83$. There is a pole at $\ell \approx 8.03$ for $\langle \rho_3 \rangle \approx 3.83$, outside of the region shown in the figure left figure, and at $\rho_3 \approx 9.37$ for $\langle \ell \rangle \approx 1.13 $ in the right figure. The overall scale of the potential has to be set by $\langle \Sigma \rangle$.}
\label{Fig_RealizationPotentialDilatonandrho3}
\end{center}
\end{figure}

\begin{figure}[ht]
\begin{center}
\includegraphics[scale=0.6]{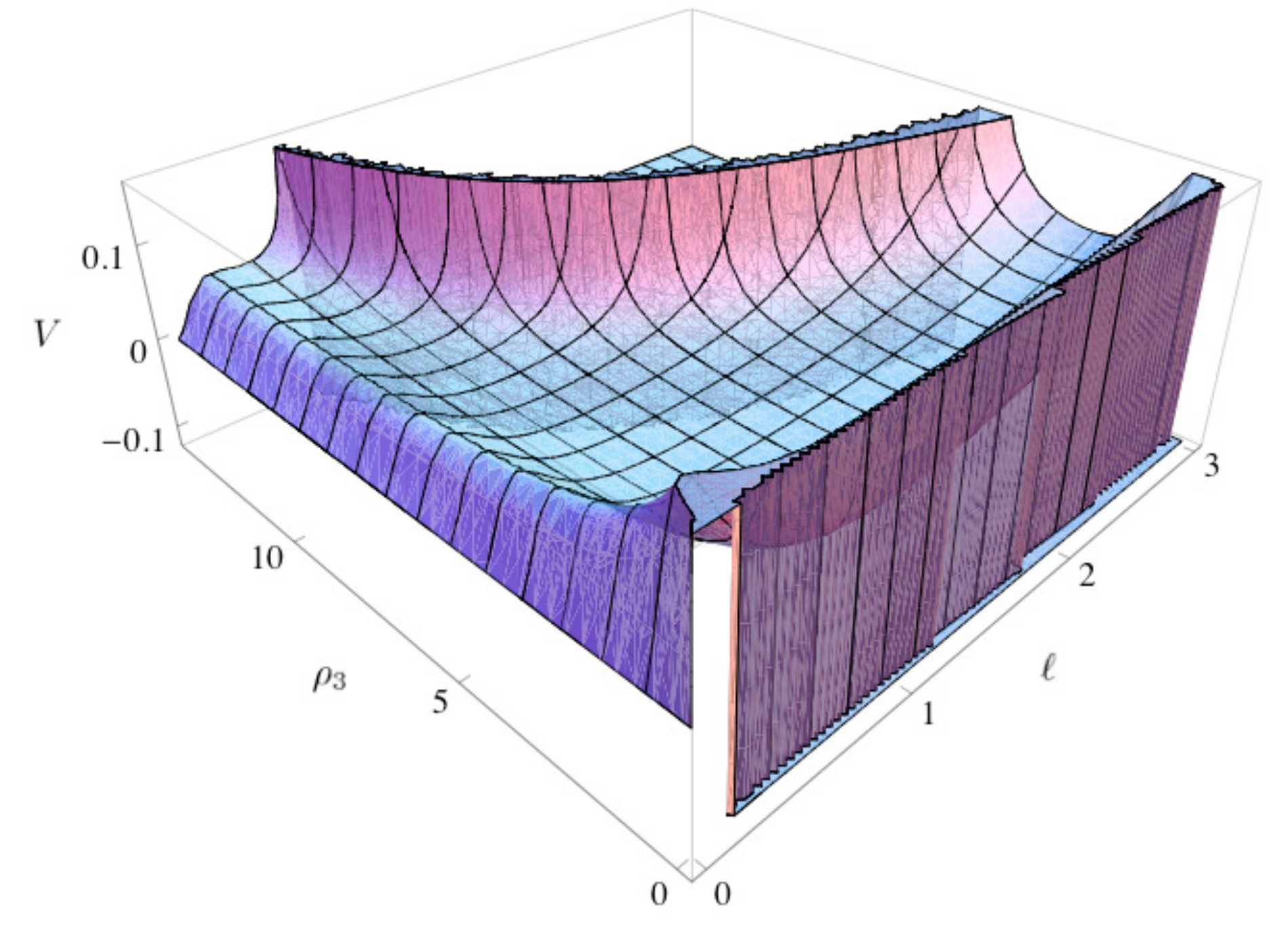}
\caption{Dependence of the potential Eq.~\eqref{Eq_RealizationPotentialDilatonrho3} on $\ell$ and $\rho_3$. For the example values $n = 1$, $q = - \frac{1}{2}$, $\gamma = \frac{10}{8 \pi^2} \approx 0.13$, $\beta_3 = \frac{30}{8 \pi^2} \approx 0.38$, $A = - 0.7$, $B = 20$ and $a = 1$, there is a minimum at $\langle \ell \rangle \approx 1.13 $ with $g^2 \approx 0.62$ and $\langle \rho_3 \rangle \approx 3.83$. The overall scale of the potential has to be set by $\langle \Sigma \rangle$.}
\label{Fig_RealizationPotentialDilatonrho3}
\end{center}
\end{figure}

The typical scale of the gaugino condensate is around $\sim 10^{11} \,\, \text{GeV}$ and thus it is negligible during inflation \cite{Gaillard:1998xx} (it is however crucial for stabilizing the dilaton and the pattern of  supersymmetry breaking after inflation). Since the setup has the properties $W \simeq 0$ and only $W_X \neq 0$, we need to stabilize the dilaton solely by the F-term of $X$ in combination with non-perturbative corrections to the K\"ahler potential, cf. Sect.~\ref{Sect_NonPerturbativeCorrections}. Therefore, we expect that the dependence of the scalar potential on the dilaton $\ell$ and $\rho_3$ as defined in Eq.~\eqref{Eq_HeisenbergInvariantCombination} can be parametrized as follows
\begin{equation}
  \label{Eq_RealizationPotentialDilatonrho3}
  V \propto \frac{ \rho_{3}^{q} \, \ell^{\, n} \, e^{g(\ell)}}{1 + \ell \gamma + \ell \beta_{3} (\log \rho_3 - \frac{\pi}{6} \rho_3 + \lambda \sum_{a} \lvert \Phi_a \rvert^2)} \, .
\end{equation}
Recall that the F-term of $X$ has a moduli dependence encoded in $\lvert a \rvert^2 \lvert \langle \Sigma \rangle \rvert^4$ and also $K_{X \bar{X}}$ depends on the moduli, cf. Eq.~\eqref{Eq_GeneralFtermpotential}. If for example $\langle \Sigma \rangle$ is independent of $\ell$, we have $n = 1$, while otherwise $n$ can be either enhanced or reduced, but it is always an integer number. Similarly, $q$ is some model-dependent rational number typically $\geq -1$. Whether $\ell$ and $\rho_3$ are stabilized or not depends on the interplay of various parameters. We have depicted the potential (in arbitrary units) as a function of $\ell$ and $\rho_3$ in Figs.~\ref{Fig_RealizationPotentialDilatonandrho3} and \ref{Fig_RealizationPotentialDilatonrho3} for an illustrative choice of parameters, which demonstrates that one can indeed stabilize $\ell$ and $\rho_3$. 

Similar to \cite{Gaillard:1998xx,Cai:1999aj}, if $n \leq 1$ the dilaton can be stabilized during inflation at $\langle \ell \rangle \sim \mathcal{O}(1)$ and with reasonable values for the gauge coupling $g$.\footnote{Note that as in \cite{Gaillard:1998xx,Cai:1999aj} at least one field contained in $\Sigma$, which collectively denotes a product of fields, has to receive an expectation value through an F-term such that the net dilaton dependence in the scalar potential satisfies $n \leq 1$.} By analogy to the discussion in Sect.~\ref{Sect_GeneralModuli}, we expect a minimum for $\rho_3$ if $q < 0$ and $\beta_3 > 0$, which indeed occurs. Interestingly, assuming $\gamma$ to be negligible, the values of $\ell$ and $\rho_3$ at their minima appear to be parametrically related by $\langle \rho_3 \rangle \sim (\beta_3 \langle \ell \rangle)^{-1}$, up to a numerical factor which is roughly $\mathcal{O}(1)$. Hence, a minimum at rather large values of $\rho_3$ requires $\beta_3 \langle \ell \rangle < 1$ and since $\beta_3$ is related to the beta function coefficient of an $\mathcal{N} = 2$ theory by $\beta_3 = b^{\, \mathcal{N}=2} / 8 \pi^2$, this can indeed be fulfilled if $\langle \ell \rangle \sim \mathcal{O}(1)$. This requirement is important, because we expect the Heisenberg symmetry to be preserved only in the large radius limit. Both $\ell$ and $\rho_3$ can be stabilized at masses $\sim H_{inf}$. Note that analogous to the situation in Sect.~\ref{Sect_GeneralModuli}, the potential has poles along a line in the $\ell$ and $\rho_3$ plane.

Concerning the pole in Fig.~\ref{Fig_RealizationPotentialDilatonandrho3}, the same discussion as under Fig.~\ref{Fig_GeneralPotentialrho3} in Sect.~\ref{Sect_GeneralModuli} applies.

\begin{figure}[ht]
\begin{center}
\centerline{
\includegraphics[scale=0.67]{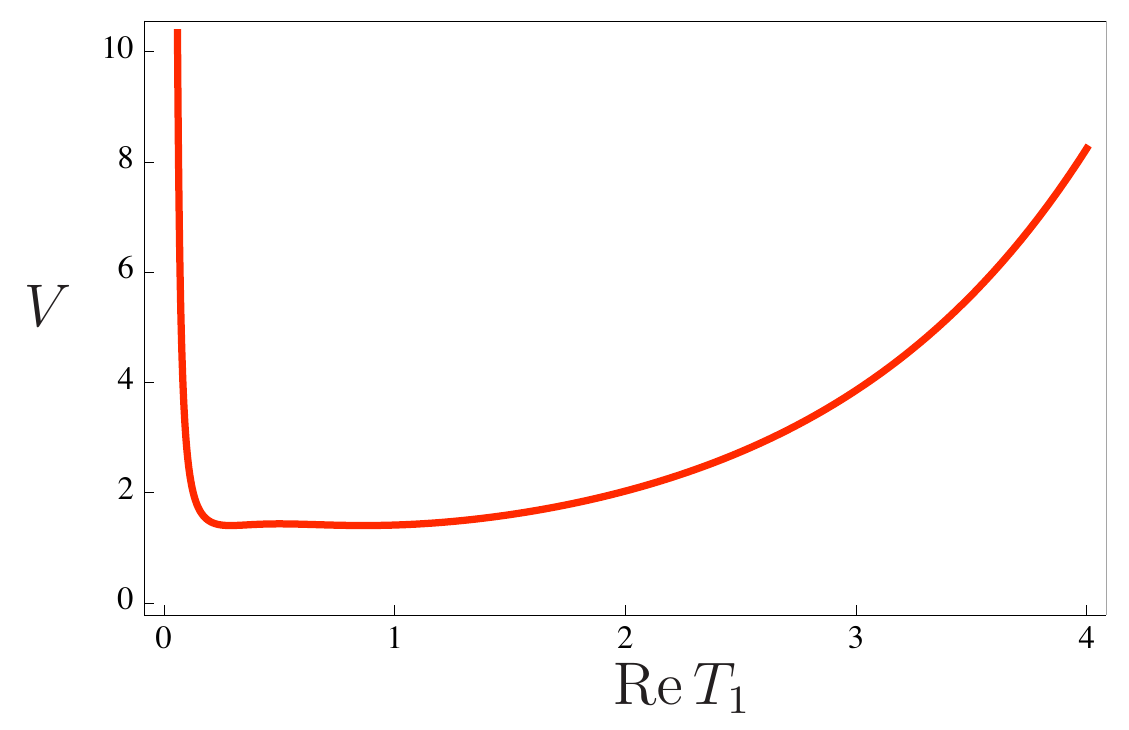}
\hspace*{.4cm}
\includegraphics[scale=0.67]{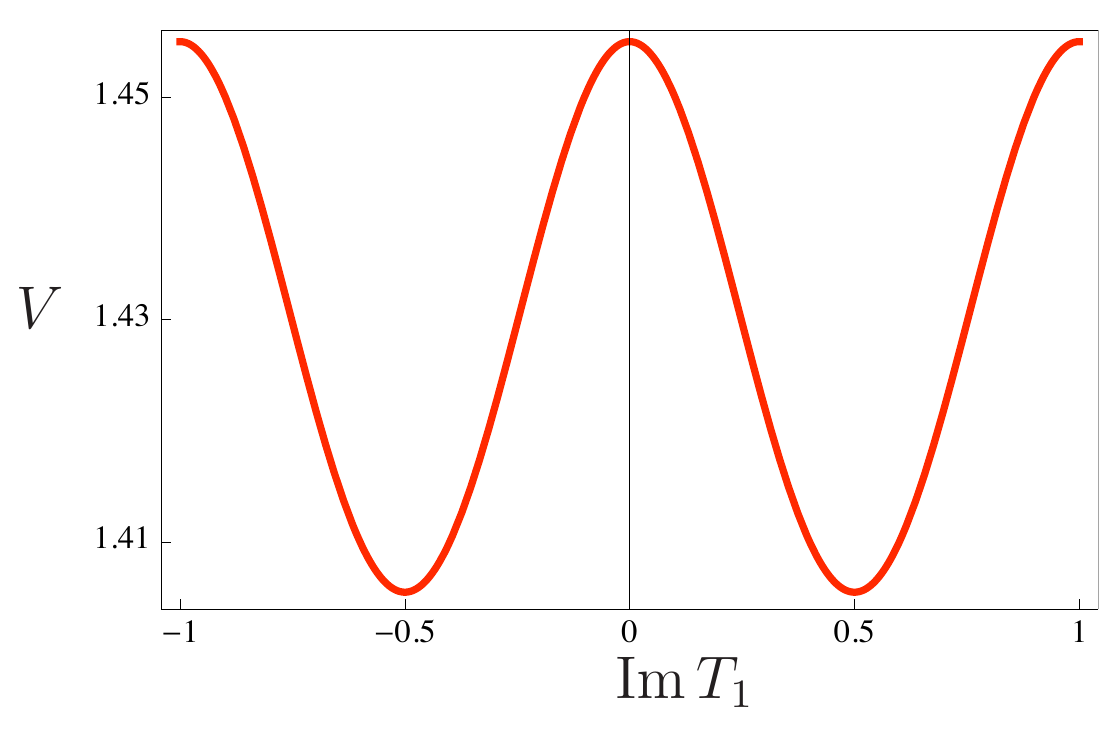}
}
\caption{Form of the potential Eq.~\eqref{Eq_RealizationPotentialT1T2} for the example values $T_i$ for $p_i = -1$, which yields a minimum at $T_i = e^{i \pi /6}$, i.e. $\text{Re} \, T_i \approx 0.87$ and $\text{Im} \, T_i = 0.5$. The overall scale of the potential has to be set by $\langle \Sigma \rangle$.}
\label{Fig_RealizationPotentialT1}
\end{center}
\end{figure}

So far, the dilaton $\ell$ and $\rho_3$ can be stabilized during inflation, but we still need to show that the remaining K\"ahler moduli $T_1$ and $T_2$ can also be fixed during inflation. They are defined as in Eq.~\eqref{Eq_ModuliDefinition} since we assume that the untwisted matter fields associated to them have negligible expectation values during inflation. The dependence of the scalar potential on $T_1$ and $T_2$ is typically of the form \cite{Copeland:1994vg}
\begin{equation}
  \label{Eq_RealizationPotentialT1T2}
  V \propto \left[ (T_i + \bar{T}_i) \, \lvert \eta(T_i) \rvert^{4} \right]^{- p_i} \, ,
\end{equation}
for some (in general rational) model-dependent numbers $p_i$. For this form of the potential, if $p_i > 0$, the $T_i$ get stabilized at the self-dual value $T_i = e^{i \pi / 6}$ with masses $\sim H_{inf}$. Note that the $\eta$-function also provides a potential for $\text{Im} \, T_i$. The stabilization of $\text{Re} \, T_i$ can be also understood from the simple example of Sect.~\ref{Sect_GeneralModuli}: for $\text{Re} \, T_i \gtrsim 1$, we can approximate $\eta(T_i) \sim \exp( - \frac{\pi}{12} T_i)$. Thus, we expect that even if $T_i + \bar{T}_i$ and $\lvert \eta(T_i) \rvert^{4}$ do not enter with the same power into the scalar potential, we get a minimum at $\text{Re} \, T_i \sim \mathcal{O}(1)$ as long as both powers are negative. Fig.~\ref{Fig_RealizationPotentialT1} shows a plot of the $T_i$ dependence of the potential Eq.~\eqref{Eq_RealizationPotentialT1T2} for a sample choice of parameters (in arbitrary units).

In the light of the considerations above, we may now assume that all moduli are stabilized with masses $\sim H_{\text{inf}}$ and regard them as effectively constant in the following.

Due to the $\mathcal{O}(1)$ values for the moduli, one might worry about higher string-loop and $\alpha^{\prime}$-corrections. The string-loop counting parameter is $\ell / 8 \pi^2$ and thus it can be sufficiently small even if $\langle \ell \rangle \sim \mathcal{O}(1)$. The issue of $\alpha^{\prime}$-corrections is a difficult question in the context of orbifolds. If they respect the Heisenberg symmetry to a sufficient amount (similar to our assumption for the string-loop corrections discussed in Sect.~\ref{Sect_LoopCorrectionsKaehler}), they will not affect the flatness of the inflaton potential, but their precise form in principle can affect e.g. moduli stabilization. At the orbifold point, i.e. at a point where the expectation values of all matter fields (in particular those of the twisted fields) vanish, one has a description in terms of an exact CFT. In the presence of an anomalous $U(1)_A$, however, some fields must acquire expectation values to cancel the FI-term, as is the case in all phenomenologically interesting orbifold models found so far. If twisted fields acquire expectation values, some (or all) of the orbifold singularities get resolved or `blown-up'. Further investigations of this issue are required, e.g. along the lines of \cite{Nibbelink:2010wm} by using gauged linear sigma models.

\subsection{D-flatness conditions}
\label{Sect_RealizationDflatness}

Before we discuss the inflaton potential, we have a few comments on constraints from D-flatness, i.e. from the requirement that the D-term potential vanishes. By assumption, the combination of untwisted matter fields $f(\Phi_a)$ in the superpotential is a gauge invariant product, such that the constraints from D-flatness are satisfied. In this way, potentially dangerous corrections to the inflation potential are avoided. We assume that $f(\Phi_a)$ carries a zero net charge under the anomalous $U(1)_A$ used to generate expectation values in Sect.~\ref{Sect_AnomalousU1andVEVs}.

Considering for simplicity a simple non-anomalous gauge group $G$, the contribution to the D-term potential is given by
\begin{equation}
  \label{Eq_RealizationDtermpotential}
  V_D = \frac{g^{2}}{2} \left( \sum_{\alpha, n} K_{\alpha} \bar{\Phi}_{\alpha} \mathcal{T}_{(\alpha)}^{\, n} \Phi_{\alpha}  \right)^2 \, ,
\end{equation}
where $\alpha$ runs over all matter fields (both untwisted and twisted) charged under the gauge group $G$, $n$ labels the generators $\mathcal{T}_{(\alpha)}^{\, n}$, which need to be taken in the appropriate representation acting on $\Phi_{\alpha}$, and $K_{\alpha}$ is the derivative of the K\"ahler potential with respect to $\Phi_{\alpha}$. In general, this is a function of the K\"ahler moduli, but since they acquire large masses as discussed above, they quickly settle to their minimum and thus the $K_{\alpha}$ are effectively constant.

Since the waterfall fields $H^{\pm}$ (which are in conjugate representations of the gauge group) are kept at zero during inflation, we focus on the $\Phi_a$. Imposing D-flatness, i.e. $V_D = 0$ in Eq.~\eqref{Eq_RealizationDtermpotential}, requires the expectation values of the $\Phi_{a}$ to fulfill certain relations. For example, in the simplest case of a $U(1)$ with $f(\Phi_a) = \Phi^+ \Phi^-$ we would have to satisfy $\lvert \Phi^+ \rvert = \lvert \Phi^- \rvert$. There are also other possibilities, e.g. the cases discussed in \cite{Antusch:2010va}, where inflation proceeds along the sneutrino direction in a Pati-Salam and $SO(10)$ supersymmetric GUT, which is one of several D-flat directions.

Note also that due to the expectation values for certain components of the $\Phi_a$ the gauge group gets broken down to a subgroup $G^{\prime} \subset G$ and the inflaton is a gauge singlet with respect to $G^{\prime}$.

\subsection{A slope for the inflaton and the hybrid mechanism}
\label{Sect_RealizationSlope}

So far, we have explained how to stabilized the moduli and considered the constraints from D-flatness in our setup. Now we discuss the form of the inflaton potential, in particular what sources can generate a slope for the inflaton direction. Inflation ends via the hybrid mechanism: once the inflaton reaches a critical value one of the waterfall fields becomes tachyonic, thereby ending inflation by a phase transition. We will not discuss this latter phase in detail and instead restrict ourselves to argue why the slope can be small.

There is already a source for a slope of the inflaton in the potential Eq.~\eqref{Eq_RealizationPotentialDilatonrho3}, namely the $\lambda \lvert \Phi_a \rvert^2$ piece, which we use to parametrize the amount of breaking of the Heisenberg symmetry by string loop corrections. With the moduli at their minimum (and before considering further corrections), the inflaton potential has the form
\begin{equation}
  \label{Eq_RealizationSlopeTreeLevel1}
  V \simeq \frac{V_0}{1 + \langle \ell \rangle \gamma + \langle \ell \rangle \beta_3 ( \log \langle \rho_3 \rangle - \frac{\pi}{6} \langle \rho_3 \rangle + \lambda \sum_{a} \lvert \phi_a \rvert^2)} \, ,
\end{equation}
where $V_0$ depends on the expectation values of $T_1$, $T_2$, $\rho_3$ and $\ell$ as well as $\langle \Sigma \rangle$. The kinetic terms of the $\phi_a$ are $\rho_{3}^{-1} \lvert \partial_{\mu} \phi_a \rvert^2$. To trigger the phase transition, the inflaton has to roll towards $\phi_a = 0$ and if the $\lambda$ term dominates the slope this requires $\lambda < 0$ since $\beta_3 > 0$, cf. Sect.~\ref{Sect_RealizationModuliStabilization}. Expanding around $\phi_a \simeq 0$ and canonically normalizing, we can estimate the contribution to the slow-roll parameter\footnote{The slow-roll parameter $\eta$ should not be confused with the Dedekind $\eta$-function $\eta(T)$ introduced above.} $\eta$:
\begin{equation}
  \label{Eq_RealizationSlopeTreeLevel2}
  \lvert \eta \rvert \sim \lvert \lambda \rvert \, ,
\end{equation}
where we used $\langle \rho_3 \rangle \sim (\beta_3 \langle \ell \rangle)^{-1}$ and $\beta_3 \langle \ell \rangle \lesssim 1$. Since slow roll inflation occurs if $\lvert \eta \rvert \ll 1$, we have to require that $\lvert \lambda \rvert \ll 1$. If the Heisenberg symmetry is broken only by non-perturbative effects such that $\lambda \sim e^{- \pi \rho_3}$ (see Sect.~\ref{Sect_LoopCorrectionsKaehler}), this condition can be fulfilled with $\langle \rho_3 \rangle$ sufficiently larger than $1$. The discussion of moduli stabilization in Sect.~\ref{Sect_RealizationModuliStabilization} implies that this indeed could be achieved.

Recalling the list of sources for a slope of Sect.~\ref{Sect_GeneralSlope}, the above contribution to the slope is a weak violation of the Heisenberg symmetry in the K\"ahler potential. In addition, if $\lambda$ is exponentially small for large $\rho_3$, other effects might also contribute significantly to the slope. One example is the slope induced by the symmetry violating term $f(\Phi_a) H^+ H^-$ in the superpotential, which enters into the one-loop Coleman-Weinberg potential. A further source is through a (small) violation of the conditions $W \simeq W_n \simeq 0$ for all $n \neq X$. Actually, such a violation is necessary since $W_X$ should be driven to zero in the waterfall phase transition and we have to stabilize the moduli also afterwards. Recently, a moduli stabilization scheme for heterotic orbifolds was proposed \cite{Dundee:2010sb}, which included the possibility that $\langle W \rangle \neq 0$ but parametrically small due to the breaking of an approximate R-symmetry at a high order in the superpotential \cite{Kappl:2008ie}. Here, such a term would induce a slope for the inflaton due to a parametrically small violation of the condition $W \simeq 0$.

To summarize, we have outlined a possibility to provide a sufficiently flat inflaton potential in combination with a mechanism to end inflation. It seems plausible that the above scenario can occur in heterotic orbifold compactifications, even though it remains to be checked whether our requirements and assumptions are fulfilled in an explicit and phenomenologically interesting model.

\section{Conclusions and outlook}
\label{Sect_Conclusions}

In this work, we have constructed a framework which is promising for realising inflation in the untwisted matter sector of heterotic orbifold compactifications. For this purpose, we have described in Sect.~\ref{Sect_GeneralModel} a class of supergravity models which is a generalization of the inflation models in \cite{Antusch:2008pn}. It is based on two ingredients necessary to solve the $\eta$-problem: a tribrid structure and a Heisenberg symmetry. The tribrid structure assigns three `tasks' to three different fields: a field $X$ provides the vacuum energy, a field $\Phi$ plays the role of the inflaton, i.e. the clock measuring when inflation ends, and a waterfall field $H$ allows inflation to end. The field $H$ has a $\Phi$-dependent mass and once $\Phi$ reaches a critical value a phase transition is triggered, which ends inflation. The superpotential during inflation satisfies $W \simeq W_n \simeq 0$ except for $n = X$, and together with the Heisenberg symmetry this protects the inflaton from dangerous corrections at tree-level. The Heisenberg symmetry allowed us to identify the inflaton $\Phi$ with a combination of gauge non-singlet matter fields. This makes such scenarios for inflation particularly appealing since one can relate particle physics models and models of inflation. A recent explicit realisation of this idea in the context of supersymmetric GUTs can be found in \cite{Antusch:2010va}.

Some features of the above class of models are typical for orbifold compactifications of the heterotic string, in particular the condition $W \simeq 0$ and the Heisenberg symmetry, which appears in the tree-level K\"ahler potential of the untwisted matter fields. Moreover, there has been a lot of progress within the last years in order to realise the MSSM in a certain class of heterotic orbifold compactifications \cite{Buchmuller:2005jr}. Very recently, \cite{Parameswaran:2010ec} attempted to find explicit metastable de Sitter vacua in this class of orbifolds, but the search has proven to be difficult.

Our aim was to realise inflation in the matter sector and with this motivation we have discussed whether the considered supergravity setup can be embedded in heterotic orbifolds. We have outlined in Sect.~\ref{Sect_Realization} under which conditions this is possible:

\begin{itemize}

  \item There exists a (tree-level) D-flat and F-flat direction of untwisted matter fields in a torus with fixed complex structure modulus.
  
  \item The relevant part of the superpotential has tribrid structure as defined in Eq.~\eqref{Eq_GeneralSuperpotential}. 
  
  \item There are suitable expectation values for the fields collectively denoted by $\langle \Sigma \rangle$, cf. Eq.~\eqref{Eq_GeneralFtermpotential}.
  
  \item The K\"ahler potential of $X$ has a moduli-dependence which respects the Heisenberg symmetry at large radius and leads to the stabilization of $\rho_3$ as defined in Eq.~\eqref{Eq_GeneralInvariantrho3}.
  
  \item The dilaton $\ell$ can be stabilized by non-perturbative corrections to the K\"ahler potential.

\end{itemize}

An important open question is whether the necessary structure of the superpotential and suitable expectation values can be realised in an explicit orbifold compactification. Also the existence of a D-flat and F-flat direction of untwisted matter fields in a torus with fixed complex structure modulus has to be verified. However, it seems plausible that these conditions can be fulfilled. It is then also particularly interesting if there is an overlap with models where spectra close to the MSSM can be achieved.

As usual in string theory models of inflation, moduli stabilization is very important but also challenging: some moduli can only be stabilized by non-perturbative effects and the large vacuum energy during inflation can lead to decompactification \cite{Kallosh:2004yh}. In the case of the Heisenberg symmetry, this is even more severe since the symmetry combines the inflaton with one of the moduli. Hence, we needed a way to stabilize the modulus without giving a large mass of the order $\sim H_{\text{inf}}$ to the inflaton. We proposed a way to achieve this based on an ansatz for the string loop corrections to the K\"ahler metric of the field $X$, which we assumed to live in a twisted sector, more precisely an $\mathcal{N} = 2$ twisted sector. In the presence of sectors with $\mathcal{N} = 2$ supersymmetry, there are known moduli-dependent threshold corrections to the matter K\"ahler metrics of untwisted matter fields \cite{Antoniadis:1992pm}.  As a working hypothesis, we assumed that these corrections have the same form for the twisted matter field K\"ahler metrics and break the Heisenberg symmetry only by terms which are exponentially suppressed in the large radius limit. It was shown that if this assumption is combined with non-perturbative corrections to the K\"ahler potential and a suitable dependence of the F-term of $X$ on the remaining K\"ahler moduli one can stabilize the bulk moduli with masses $m \gtrsim H_{\text{inf}}$. Unfortunately, the K\"ahler potential for twisted matter fields is typically only known at tree-level and to quadratic order in the fields. We have proposed a form which is interesting from the point of view of inflation and seems reasonable. However, it remains to be seen whether our hypothesis can be fulfilled in an explicit and phenomenologically interesting model.

Assuming all requirements for our setup on the structure of the K\"ahler and superpotential can be fulfilled, one needs a concrete model at hand in order to make predictions for observables, for example the scalar spectral index $n_s$. One has to determine the field content and the superpotential to fix the moduli dependence of the scalar potential. In addition, one needs to compute all the relevant parameters from the underlying string theory, which is however quite difficult, especially for the non-perturbative corrections to the K\"ahler potential. 

A further important task is to gain a more detailed understanding of the origin and the fate of the Heisenberg symmetry within string theory. Its 10d realization involves a shift of the NS 2-form $B_{M N}$ and therefore we expect a connection with T-duality which needs to be clarified.

In summary, we have proposed a framework for realising inflation in the matter sector of heterotic orbifold compactifications. Scenarios with a matter field as the inflaton are phenomenologically attractive since they relate models of inflation and particle physics. Our present work should be viewed as a first step towards this goal and we have discussed the conditions which have to be fulfilled in an explicit heterotic orbifold model.

\subsubsection*{Acknowledgements}

We are grateful to Stefan Groot Nibbelink, Michael Haack, Dieter L\"ust, Stephan Stieberger, Alexander Westphal and Marco Zagermann for discussions and comments.

This work has been supported by the Cluster of Excellence `Origin and Structure of the Universe'. K. D. is supported by the German Science Foundation (DFG) within the Collaborative Research Center 676 `Particles, Strings and the Early Universe'.


\begin{thebibliography}{10}

\bibitem{Lyth:2009zz}
  For textbook reviews on inflation see: D.~H.~Lyth and A.~R.~Liddle,
  ``The Primordial Density Perturbation: Cosmology, Inflation and the Origin of
  Structure,''
  {\it  Cambridge, UK: Cambridge Univ. Pr. (2009) 497 p}; 
  A.~D.~Linde, ``Particle Physics and Inflationary Cosmology,''
  [arXiv:hep-th/0503203]; 
  V.~Mukhanov, ``Physical Foundations of Cosmology,''
  {\it  Cambridge, UK: Univ. Pr. (2005) 421 p}.

\bibitem{Copeland:1994vg}
  E.~J.~Copeland, A.~R.~Liddle, D.~H.~Lyth, E.~D.~Stewart and D.~Wands,
  ``False vacuum inflation with Einstein gravity,''
  Phys.\ Rev.\  D {\bf 49}, 6410 (1994)
  [arXiv:astro-ph/9401011].
  
\bibitem{Dvali:1995mj}
  G.~R.~Dvali,
  ``Inflation versus the cosmological moduli problem,''
  arXiv:hep-ph/9503259;\\
%
  M.~Dine, L.~Randall and S.~D.~Thomas,
  ``Supersymmetry breaking in the early universe,''
  Phys.\ Rev.\ Lett.\  {\bf 75}, 398 (1995)
  [arXiv:hep-ph/9503303].
  
\bibitem{Kawasaki:2000yn}
  M.~Kawasaki, M.~Yamaguchi and T.~Yanagida,
  ``Natural chaotic inflation in supergravity,''
  Phys.\ Rev.\ Lett.\  {\bf 85}, 3572 (2000)
  [arXiv:hep-ph/0004243].
  
\bibitem{Brax:2005jv}
  P.~Brax and J.~Martin,
  ``Shift symmetry and inflation in supergravity,''
  Phys.\ Rev.\  D {\bf 72}, 023518 (2005)
  [arXiv:hep-th/0504168].
  
\bibitem{Stewart:1994ts}
  E.~D.~Stewart,
  ``Inflation, Supergravity And Superstrings,''
  Phys.\ Rev.\  D {\bf 51} (1995) 6847
  [arXiv:hep-ph/9405389].

\bibitem{Gaillard:1995az}
  M.~K.~Gaillard, H.~Murayama and K.~A.~Olive,
  ``Preserving flat directions during inflation,''
  Phys.\ Lett.\  B {\bf 355} (1995) 71
  [arXiv:hep-ph/9504307].
  
\bibitem{Antusch:2008pn}
  S.~Antusch, M.~Bastero-Gil, K.~Dutta, S.~F.~King and P.~M.~Kostka,
  ``Solving the $\eta$-Problem in Hybrid Inflation with Heisenberg Symmetry and
  Stabilized Modulus,''
  JCAP {\bf 0901} (2009) 040
  [arXiv:0808.2425 [hep-ph]].
  
\bibitem{Linde:1993cn}
  A.~D.~Linde,
  ``Hybrid inflation,''
  Phys.\ Rev.\  D {\bf 49} (1994) 748
  [arXiv:astro-ph/9307002].
  
\bibitem{Dvali:1994ms}
  G.~R.~Dvali, Q.~Shafi and R.~K.~Schaefer,
  ``Large scale structure and supersymmetric inflation without fine tuning,''
  Phys.\ Rev.\ Lett.\  {\bf 73} (1994) 1886
  [arXiv:hep-ph/9406319].

\bibitem{Antusch:2009vg}
  S.~Antusch, K.~Dutta and P.~M.~Kostka,
  ``Tribrid Inflation in Supergravity,''
  AIP Conf.\ Proc.\  {\bf 1200}, 1007 (2010)
  [arXiv:0908.1694 [hep-ph]].
  
\bibitem{Kallosh:2010xz}
  R.~Kallosh, A.~Linde and T.~Rube,
  ``General inflaton potentials in supergravity,''
  arXiv:1011.5945 [hep-th].
  
\bibitem{Antusch:2010va}
  S.~Antusch, M.~Bastero-Gil, J.~P.~Baumann, K.~Dutta, S.~F.~King and P.~M.~Kostka,
  ``Gauge Non-Singlet Inflation in SUSY GUTs,''
  JHEP {\bf 1008} (2010) 100
  [arXiv:1003.3233 [hep-ph]].
  
\bibitem{Kallosh:1995hi}
  R.~Kallosh, A.~D.~Linde, D.~A.~Linde and L.~Susskind,
  ``Gravity and global symmetries,''
  Phys.\ Rev.\  D {\bf 52} (1995) 912
  [arXiv:hep-th/9502069];\\
%
  M.~Kamionkowski and J.~March-Russell,
  ``Are textures natural?,''
  Phys.\ Rev.\ Lett.\  {\bf 69} (1992) 1485
  [arXiv:hep-th/9201063];\\
%
  M.~Kamionkowski and J.~March-Russell,
  ``Planck scale physics and the Peccei-Quinn mechanism,''
  Phys.\ Lett.\  B {\bf 282} (1992) 137
  [arXiv:hep-th/9202003];\\
%
  R.~Holman, S.~D.~H.~Hsu, T.~W.~Kephart, E.~W.~Kolb, R.~Watkins and L.~M.~Widrow,
  ``Solutions to the strong CP problem in a world with gravity,''
  Phys.\ Lett.\  B {\bf 282} (1992) 132
  [arXiv:hep-ph/9203206];\\
%
  T.~Banks and N.~Seiberg,
  ``Symmetries and Strings in Field Theory and Gravity,''
  arXiv:1011.5120 [hep-th].
  
\bibitem{Baumann:2010ys}
  D.~Baumann and D.~Green,
  ``Desensitizing Inflation from the Planck Scale,''
  JHEP {\bf 1009} (2010) 057
  [arXiv:1004.3801 [hep-th]].
  
\bibitem{Baumann:2010nu}
  D.~Baumann and D.~Green,
  ``Inflating with Baryons,''
  arXiv:1009.3032 [hep-th].
  
\bibitem{Dvali:1998pa}
  G.~R.~Dvali and S.~H.~H.~Tye,
  ``Brane inflation,''
  Phys.\ Lett.\  B {\bf 450}, 72 (1999)
  [arXiv:hep-ph/9812483];\\
%
  S.~H.~S.~Alexander,
  ``Inflation from D - anti-D brane annihilation,''
  Phys.\ Rev.\  D {\bf 65}, 023507 (2002)
  [arXiv:hep-th/0105032];\\
%
  C.~P.~Burgess, M.~Majumdar, D.~Nolte, F.~Quevedo, G.~Rajesh and R.~J.~Zhang,
  ``The Inflationary Brane-Antibrane Universe,''
  JHEP {\bf 0107}, 047 (2001)
  [arXiv:hep-th/0105204].
  
\bibitem{Kachru:2003sx}
  S.~Kachru, R.~Kallosh, A.~D.~Linde, J.~M.~Maldacena, L.~P.~McAllister and S.~P.~Trivedi,
  ``Towards inflation in string theory,''
  JCAP {\bf 0310} (2003) 013
  [arXiv:hep-th/0308055].
  
\bibitem{Berg:2004ek}
  M.~Berg, M.~Haack and B.~K\"ors,
  ``Loop corrections to volume moduli and inflation in string theory,''
  Phys.\ Rev.\  D {\bf 71} (2005) 026005
  [arXiv:hep-th/0404087]; \\
%
  M.~Berg, M.~Haack and B.~K\"ors,
  ``On the moduli dependence of nonperturbative superpotentials in brane
  inflation,''
  arXiv:hep-th/0409282;\\
%
  D.~Baumann, A.~Dymarsky, I.~R.~Klebanov, J.~M.~Maldacena, L.~P.~McAllister and A.~Murugan,
  ``On D3-brane potentials in compactifications with fluxes and wrapped
  D-branes,''
  JHEP {\bf 0611} (2006) 031
  [arXiv:hep-th/0607050];\\
%
  D.~Baumann, A.~Dymarsky, I.~R.~Klebanov, L.~McAllister and P.~J.~Steinhardt,
  ``A Delicate Universe,''
  Phys.\ Rev.\ Lett.\  {\bf 99} (2007) 141601
  [arXiv:0705.3837 [hep-th]];\\
%
  D.~Baumann, A.~Dymarsky, I.~R.~Klebanov and L.~McAllister,
  ``Towards an Explicit Model of D-brane Inflation,''
  JCAP {\bf 0801} (2008) 024
  [arXiv:0706.0360 [hep-th]];\\
%
  D.~Baumann, A.~Dymarsky, S.~Kachru, I.~R.~Klebanov and L.~McAllister,
  ``Holographic Systematics of D-brane Inflation,''
  JHEP {\bf 0903} (2009) 093
  [arXiv:0808.2811 [hep-th]];\\
%
  D.~Baumann, A.~Dymarsky, S.~Kachru, I.~R.~Klebanov and L.~McAllister,
  ``Compactification Effects in D-brane Inflation,''
  Phys.\ Rev.\ Lett.\  {\bf 104} (2010) 251602
  [arXiv:0912.4268 [hep-th]];\\
%
  D.~Baumann, A.~Dymarsky, S.~Kachru, I.~R.~Klebanov and L.~McAllister,
  ``D3-brane Potentials from Fluxes in AdS/CFT,''
  JHEP {\bf 1006} (2010) 072
  [arXiv:1001.5028 [hep-th]];\\
%
  C.~P.~Burgess, J.~M.~Cline, K.~Dasgupta and H.~Firouzjahi,
  ``Uplifting and inflation with D3 branes,''
  JHEP {\bf 0703} (2007) 027
  [arXiv:hep-th/0610320];\\
%
  A.~Krause and E.~Pajer,
  ``Chasing Brane Inflation in String-Theory,''
  JCAP {\bf 0807} (2008) 023
  [arXiv:0705.4682 [hep-th]];\\
%
  H.~Y.~Chen, L.~Y.~Hung and G.~Shiu,
  ``Inflation on an Open Racetrack,''
  JHEP {\bf 0903} (2009) 083
  [arXiv:0901.0267 [hep-th]].
  
\bibitem{Haack:2008yb}
  M.~Haack, R.~Kallosh, A.~Krause, A.~D.~Linde, D.~L\"ust and M.~Zagermann,
  ``Update of D3/D7-Brane Inflation on $K3 \times T^2/\mathbb{Z}_2$,''
  Nucl.\ Phys.\  B {\bf 806} (2009) 103
  [arXiv:0804.3961 [hep-th]];\\
%
  C.~P.~Burgess, J.~M.~Cline and M.~Postma,
  ``Axionic D3-D7 Inflation,''
  JHEP {\bf 0903} (2009) 058
  [arXiv:0811.1503 [hep-th]].

\bibitem{BlancoPillado:2004ns}
  J.~J.~Blanco-Pillado {\it et al.},
  ``Racetrack inflation,''
  JHEP {\bf 0411}, 063 (2004)
  [arXiv:hep-th/0406230];\\
%
  J.~P.~Conlon and F.~Quevedo,
  ``K\"ahler moduli inflation,''
  JHEP {\bf 0601}, 146 (2006)
  [arXiv:hep-th/0509012].
  
\bibitem{Silverstein:2008sg}
  E.~Silverstein and A.~Westphal,
  ``Monodromy in the CMB: Gravity Waves and String Inflation,''
  Phys.\ Rev.\  D {\bf 78}, 106003 (2008)
  [arXiv:0803.3085 [hep-th]].
  
\bibitem{McAllister:2007bg}
  L.~McAllister and E.~Silverstein,
  ``String Cosmology: A Review,''
  Gen.\ Rel.\ Grav.\  {\bf 40}, 565 (2008)
  [arXiv:0710.2951 [hep-th]];\\
%
  D.~Baumann and L.~McAllister,
  ``Advances in Inflation in String Theory,''
  Ann.\ Rev.\ Nucl.\ Part.\ Sci.\  {\bf 59}, 67 (2009)
  [arXiv:0901.0265 [hep-th]];\\
%
  J.~Erdmenger (Ed.),
  ``String cosmology: Modern string theory concepts from the Big Bang to cosmic
  structure,''
  {\it  Weinheim, Germany: Wiley-VCH (2009) 313 p}.
   
\bibitem{Giddings:2001yu}
  S.~B.~Giddings, S.~Kachru and J.~Polchinski,
  ``Hierarchies from fluxes in string compactifications,''
  Phys.\ Rev.\  D {\bf 66} (2002) 106006
  [arXiv:hep-th/0105097].
  
\bibitem{Kachru:2003aw}
  S.~Kachru, R.~Kallosh, A.~D.~Linde and S.~P.~Trivedi,
  ``De Sitter vacua in string theory,''
  Phys.\ Rev.\  D {\bf 68}, 046005 (2003)
  [arXiv:hep-th/0301240].
  
\bibitem{Grana:2005jc}
  M.~Gra\~{n}a,
  ``Flux compactifications in string theory: A comprehensive review,''
  Phys.\ Rept.\  {\bf 423} (2006) 91
  [arXiv:hep-th/0509003];\\
%
  M.~R.~Douglas and S.~Kachru,
  ``Flux compactification,''
  Rev.\ Mod.\ Phys.\  {\bf 79} (2007) 733
  [arXiv:hep-th/0610102];\\
%
  R.~Blumenhagen, B.~Kors, D.~L\"ust and S.~Stieberger,
  ``Four-dimensional String Compactifications with D-Branes, Orientifolds   and
  Fluxes,''
  Phys.\ Rept.\  {\bf 445} (2007) 1
  [arXiv:hep-th/0610327].
  
\bibitem{Binetruy:1987xj}
  P.~Binetruy and M.~K.~Gaillard,
  ``Noncompact Symmetries and Scalar Masses in Superstring - Inspired Models,''
  Phys.\ Lett.\  B {\bf 195}, 382 (1987).
  
\bibitem{Ellwanger:1986yh}
  U.~Ellwanger and M.~G.~Schmidt,
  ``SU(N,1) Supergravity from Superstrings,''
  Nucl.\ Phys.\  B {\bf 294} (1987) 445.
  
\bibitem{Gaillard:1998xx}
  M.~K.~Gaillard, D.~H.~Lyth and H.~Murayama,
  ``Inflation and flat directions in modular invariant superstring  effective
  theories,''
  Phys.\ Rev.\  D {\bf 58}, 123505 (1998)
  [arXiv:hep-th/9806157].
  
\bibitem{Buchmuller:2005jr}
  W.~Buchm\"uller, K.~Hamaguchi, O.~Lebedev and M.~Ratz,
  ``Supersymmetric standard model from the heterotic string,''
  Phys.\ Rev.\ Lett.\  {\bf 96} (2006) 121602
  [arXiv:hep-ph/0511035];\\
%
  W.~Buchm\"uller, K.~Hamaguchi, O.~Lebedev and M.~Ratz,
  ``Supersymmetric standard model from the heterotic string. II,''
  Nucl.\ Phys.\  B {\bf 785} (2007) 149
  [arXiv:hep-th/0606187];\\
%
  O.~Lebedev, H.~P.~Nilles, S.~Raby, S.~Ramos-Sanchez, M.~Ratz, P.~K.~S.~Vaudrevange and A.~Wingerter,
  ``A mini-landscape of exact MSSM spectra in heterotic orbifolds,''
  Phys.\ Lett.\  B {\bf 645} (2007) 88
  [arXiv:hep-th/0611095];\\
%
  O.~Lebedev, H.~P.~Nilles, S.~Raby, S.~Ramos-Sanchez, M.~Ratz, P.~K.~S.~Vaudrevange and A.~Wingerter,
  ``The Heterotic Road to the MSSM with R parity,''
  Phys.\ Rev.\  D {\bf 77} (2008) 046013
  [arXiv:0708.2691 [hep-th]];\\
%
  O.~Lebedev, H.~P.~Nilles, S.~Ramos-Sanchez, M.~Ratz and P.~K.~S.~Vaudrevange,
  ``Heterotic mini-landscape (II): completing the search for MSSM vacua in a
  $\mathbb{Z}_6$ orbifold,''
  Phys.\ Lett.\  B {\bf 668} (2008) 331
  [arXiv:0807.4384 [hep-th]].

\bibitem{Dundee:2010sb}
  B.~Dundee, S.~Raby and A.~Westphal,
  ``Moduli stabilization and SUSY breaking in heterotic orbifold string
  models,''
  arXiv:1002.1081 [hep-th].
  
\bibitem{Parameswaran:2010ec}
  S.~L.~Parameswaran, S.~Ramos-Sanchez and I.~Zavala,
  ``On Moduli Stabilisation and de Sitter Vacua in MSSM Heterotic Orbifolds,''
  arXiv:1009.3931 [hep-th].
  
\bibitem{Kallosh:2004yh}
  R.~Kallosh and A.~D.~Linde,
  ``Landscape, the scale of SUSY breaking, and inflation,''
  JHEP {\bf 0412} (2004) 004
  [arXiv:hep-th/0411011];\\
%
  J.~P.~Conlon, R.~Kallosh, A.~D.~Linde and F.~Quevedo,
  ``Volume Modulus Inflation and the Gravitino Mass Problem,''
  JCAP {\bf 0809} (2008) 011
  [arXiv:0806.0809 [hep-th]].

\bibitem{He:2010uk}
  T.~He, S.~Kachru and A.~Westphal,
  ``Gravity waves and the LHC: Towards high-scale inflation with low-energy
  SUSY,''
  JHEP {\bf 1006} (2010) 065
  [arXiv:1003.4265 [hep-th]].

\bibitem{Antoniadis:1992pm}
  I.~Antoniadis, E.~Gava, K.~S.~Narain and T.~R.~Taylor,
  ``Superstring threshold corrections to Yukawa couplings,''
  Nucl.\ Phys.\  B {\bf 407} (1993) 706
  [arXiv:hep-th/9212045].
  
\bibitem{Dine:1986zy}
  M.~Dine, N.~Seiberg, X.~G.~Wen and E.~Witten,
  ``Nonperturbative Effects on the String World Sheet,''
  Nucl.\ Phys.\  B {\bf 278} (1986) 769;\\
%
  M.~Dine, N.~Seiberg, X.~G.~Wen and E.~Witten,
  ``Nonperturbative Effects on the String World Sheet. 2,''
  Nucl.\ Phys.\  B {\bf 289} (1987) 319.
  
\bibitem{Ferrara:1991uz}
  S.~Ferrara, C.~Kounnas, D.~L\"ust and F.~Zwirner,
  ``Duality invariant partition functions and automorphic superpotentials for
  (2,2) string compactifications,''
  Nucl.\ Phys.\  B {\bf 365} (1991) 431.
  
\bibitem{Covi:2008cn}
  L.~Covi, M.~Gomez-Reino, C.~Gross, J.~Louis, G.~A.~Palma and C.~A.~Scrucca,
  ``Constraints on modular inflation in supergravity and string theory,''
  JHEP {\bf 0808} (2008) 055
  [arXiv:0805.3290 [hep-th]].
  
\bibitem{Jeannerot:2005mc}
  R.~Jeannerot and M.~Postma,
  ``Confronting hybrid inflation in supergravity with CMB data,''
  JHEP {\bf 0505} (2005) 071
  [arXiv:hep-ph/0503146];\\
%
  R.~Jeannerot and M.~Postma,
  ``Leptogenesis from reheating after inflation and cosmic string decay,''
  JCAP {\bf 0512} (2005) 006
  [arXiv:hep-ph/0507162];\\
%
  R.~A.~Battye, B.~Garbrecht and A.~Moss,
  ``Constraints on supersymmetric models of hybrid inflation,''
  JCAP {\bf 0609} (2006) 007
  [arXiv:astro-ph/0607339].

\bibitem{Dvali:1995fb}
  G.~R.~Dvali,
  ``Inflation Induced Susy Breaking And Flat Vacuum Directions,''
  Phys.\ Lett.\  B {\bf 355} (1995) 78
  [arXiv:hep-ph/9503375].
  
\bibitem{Bailin:1999nk}
  D.~Bailin and A.~Love,
  ``Orbifold compactifications of string theory,''
  Phys.\ Rept.\  {\bf 315} (1999) 285.
  
\bibitem{Witten:1985xb}
  E.~Witten,
  ``Dimensional Reduction Of Superstring Models,''
  Phys.\ Lett.\  B {\bf 155} (1985) 151;\\
%
  S.~Ferrara, C.~Kounnas and M.~Porrati,
  ``General Dimensional Reduction Of Ten-Dimensional Supergravity And
  Superstring,''
  Phys.\ Lett.\  B {\bf 181} (1986) 263;\\
%
  M.~Cvetic, J.~Louis and B.~A.~Ovrut,
  ``A String Calculation of the K\"ahler Potentials for Moduli of Z(N)
  Orbifolds,''
  Phys.\ Lett.\  B {\bf 206} (1988) 227;\\
%
  M.~Cvetic, J.~Molera and B.~A.~Ovrut,
  ``K\"ahler Potentials For Matter Scalars And Moduli Of Z(N) Orbifolds,''
  Phys.\ Rev.\  D {\bf 40} (1989) 1140.

\bibitem{LopesCardoso:1994is}
  G.~Lopes Cardoso, D.~L\"ust and T.~Mohaupt,
  ``Moduli spaces and target space duality symmetries in (0,2) Z(N) orbifold
  theories with continuous Wilson lines,''
  Nucl.\ Phys.\  B {\bf 432} (1994) 68
  [arXiv:hep-th/9405002].
  
\bibitem{Dixon:1989fj}
  L.~J.~Dixon, V.~Kaplunovsky and J.~Louis,
  ``On Effective Field Theories Describing (2,2) Vacua of the Heterotic
  String,''
  Nucl.\ Phys.\  B {\bf 329} (1990) 27.
  
\bibitem{Burgess:1995kp}
  C.~P.~Burgess, J.~P.~Derendinger, F.~Quevedo and M.~Quiros,
  ``Gaugino condensates and chiral linear duality: An Effective Lagrangian
  analysis,''
  Phys.\ Lett.\  B {\bf 348} (1995) 428
  [arXiv:hep-th/9501065].

\bibitem{Shapere:1988zv}
  A.~D.~Shapere and F.~Wilczek,
  ``Selfdual Models with Theta Terms,''
  Nucl.\ Phys.\  B {\bf 320} (1989) 669;\\
%
  S.~Ferrara, D.~L\"ust, A.~D.~Shapere and S.~Theisen,
  ``Modular Invariance in Supersymmetric Field Theories,''
  Phys.\ Lett.\  B {\bf 225} (1989) 363.

\bibitem{Ibanez:1992hc}
  L.~E.~Ibanez and D.~L\"ust,
  ``Duality anomaly cancellation, minimal string unification and the effective
  low-energy Lagrangian of 4-D strings,''
  Nucl.\ Phys.\  B {\bf 382} (1992) 305
  [arXiv:hep-th/9202046].
  
\bibitem{Derendinger:1991hq}
  J.~P.~Derendinger, S.~Ferrara, C.~Kounnas and F.~Zwirner,
  ``On loop corrections to string effective field theories: Field dependent
  gauge couplings and sigma model anomalies,''
  Nucl.\ Phys.\  B {\bf 372} (1992) 145.
  
\bibitem{Derendinger:1991kr}
  J.~P.~Derendinger, S.~Ferrara, C.~Kounnas and F.~Zwirner,
  ``All loop gauge couplings from anomaly cancellation in string effective
  theories,''
  Phys.\ Lett.\  B {\bf 271} (1991) 307;\\
%
  L.~J.~Dixon, V.~Kaplunovsky and J.~Louis,
  ``Moduli dependence of string loop corrections to gauge coupling constants,''
  Nucl.\ Phys.\  B {\bf 355} (1991) 649.

\bibitem{Lust:1991yi}
  D.~L\"ust and C.~Munoz,
  ``Duality invariant gaugino condensation and one loop corrected K\"ahler
  potentials in string theory,''
  Phys.\ Lett.\  B {\bf 279} (1992) 272
  [arXiv:hep-th/9201047].

\bibitem{Kaplunovsky:1995jw}
  V.~Kaplunovsky and J.~Louis,
  ``On Gauge couplings in string theory,''
  Nucl.\ Phys.\  B {\bf 444} (1995) 191
  [arXiv:hep-th/9502077].

\bibitem{LopesCardoso:1991zt}
  G.~Lopes Cardoso and B.~A.~Ovrut,
  ``A Green-Schwarz mechanism for D = 4, N=1 supergravity anomalies,''
  Nucl.\ Phys.\  B {\bf 369} (1992) 351;\\
%
  G.~Lopes Cardoso and B.~A.~Ovrut,
  ``Coordinate and Kahler sigma model anomalies and their cancellation in
  string effective field theories,''
  Nucl.\ Phys.\  B {\bf 392} (1993) 315
  [arXiv:hep-th/9205009].
  
\bibitem{Kain:2006nx}
  B.~Kain,
  ``'Semi-Realistic' F-term Inflation Model Building in Supergravity,''
  Nucl.\ Phys.\  B {\bf 800} (2008) 270
  [arXiv:hep-ph/0608279].

\bibitem{Binetruy:1996xja}
  P.~Binetruy, M.~K.~Gaillard and Y.~Y.~Wu,
  ``Dilaton Stabilization in the Context of Dynamical Supersymmetry Breaking
  through Gaugino Condensation,''
  Nucl.\ Phys.\  B {\bf 481}, 109 (1996)
  [arXiv:hep-th/9605170];\\
%
  P.~Binetruy, M.~K.~Gaillard and Y.~Y.~Wu,
  ``Modular invariant formulation of multi-gaugino and matter  condensation,''
  Nucl.\ Phys.\  B {\bf 493}, 27 (1997)
  [arXiv:hep-th/9611149];\\
%
  P.~Binetruy, M.~K.~Gaillard and Y.~Y.~Wu,
  ``Supersymmetry breaking and weakly vs. strongly coupled string theory,''
  Phys.\ Lett.\  B {\bf 412}, 288 (1997)
  [arXiv:hep-th/9702105];\\
%
  M.~K.~Gaillard and B.~D.~Nelson,
  ``K\"ahler stabilized, modular invariant heterotic string models,''
  Int.\ J.\ Mod.\ Phys.\  A {\bf 22} (2007) 1451
  [arXiv:hep-th/0703227].
  
\bibitem{Banks:1994sg}
  T.~Banks and M.~Dine,
  ``Coping With Strongly Coupled String Theory,''
  Phys.\ Rev.\  D {\bf 50} (1994) 7454
  [arXiv:hep-th/9406132].

\bibitem{Shenker:1990uf}
  S.~H.~Shenker,
  ``The Strength of nonperturbative effects in string theory,'' \\ {\it "Presented at the Cargese Workshop on Random Surfaces,
                  Quantum Gravity and Strings, Cargese, France, May 28 - Jun
                  1, 1990"}

\bibitem{Polchinski:1994fq}
  J.~Polchinski,
  ``Combinatorics of boundaries in string theory,''
  Phys.\ Rev.\  D {\bf 50} (1994) 6041
  [arXiv:hep-th/9407031];\\
%
  E.~Silverstein,
  ``Duality, compactification, and $e^{-1 / \lambda}$ effects in the heterotic
  string theory,''
  Phys.\ Lett.\  B {\bf 396} (1997) 91
  [arXiv:hep-th/9611195];\\
%
  I.~Antoniadis, B.~Pioline and T.~R.~Taylor,
  ``Calculable $e^{-1 / \lambda}$ effects,''
  Nucl.\ Phys.\  B {\bf 512} (1998) 61
  [arXiv:hep-th/9707222].
  
\bibitem{Font:1990nt}
  A.~Font, L.~E.~Ibanez, D.~L\"ust and F.~Quevedo,
  ``Supersymmetry Breaking from Duality Invariant Gaugino Condensation,''
  Phys.\ Lett.\  B {\bf 245} (1990) 401;\\
%
  J.~A.~Casas, Z.~Lalak, C.~Munoz and G.~G.~Ross,
  ``Hierarchical Supersymmetry Breaking and Dynamical Determination of
  Compactification Parameters by Nonperturbative Effects,''
  Nucl.\ Phys.\  B {\bf 347} (1990) 243;\\
%
  B.~de Carlos, J.~A.~Casas and C.~Munoz,
  ``Supersymmetry breaking and determination of the unification gauge coupling
  constant in string theories,''
  Nucl.\ Phys.\  B {\bf 399}, 623 (1993)
  [arXiv:hep-th/9204012].

\bibitem{Casas:1996zi}
  J.~A.~Casas,
  ``The generalized dilaton supersymmetry breaking scenario,''
  Phys.\ Lett.\  B {\bf 384} (1996) 103
  [arXiv:hep-th/9605180];\\
%
  T.~Barreiro, B.~de Carlos and E.~J.~Copeland,
  ``On non-perturbative corrections to the K\"ahler potential,''
  Phys.\ Rev.\  D {\bf 57} (1998) 7354
  [arXiv:hep-ph/9712443].
  
\bibitem{Cai:1999aj}
  M.~J.~Cai and M.~K.~Gaillard,
  ``Comment on 'Inflation and flat directions in modular invariant  superstring
  effective theories',''
  Phys.\ Rev.\  D {\bf 62} (2000) 047901
  [arXiv:hep-ph/9910478].
  
\bibitem{Nibbelink:2010wm}
  S.~Groot Nibbelink,
  ``Heterotic orbifold resolutions as (2,0) gauged linear sigma models,''
  arXiv:1012.3350 [hep-th].
  
\bibitem{Kappl:2008ie}
  R.~Kappl, H.~P.~Nilles, S.~Ramos-Sanchez, M.~Ratz, K.~Schmidt-Hoberg and P.~K.~S.~Vaudrevange,
  ``Large hierarchies from approximate R symmetries,''
  Phys.\ Rev.\ Lett.\  {\bf 102} (2009) 121602
  [arXiv:0812.2120 [hep-th]].
  
\end{thebibliography}
\end{document}